\documentclass[12pt]{article}
\usepackage{graphicx}
\usepackage{enumerate}
\usepackage{float}
\usepackage{latexsym,amsmath,mathrsfs,bm,amssymb,color,hiroshima}
\newtheorem{theorem}{Theorem}[section]
\newtheorem{proposition}[theorem]{Proposition}
\newtheorem{lemma}[theorem]{Lemma}
\newtheorem{corollary}[theorem]{Corollary}
\newtheorem{definition}[theorem]{Definition}
\newtheorem{example}[theorem]{Example}
\newtheorem{remark}[theorem]{Remark}

\makeatletter
\@addtoreset{equation}{section}
\makeatother
\def\theequation{\arabic{section}.\arabic{equation}}
\renewcommand{\ker}{\mathop{\mathrm{Ker}}}
\newcommand{\Log}{\mathop{\mathrm{Log}}}

\begin{document}
\title{\sc 
Conjugate operators of one-dimensional harmonic oscillator}
\author{Fumio Hiroshima\footnote{Faculty of Mathematics, Kyushu University, corresponding author,hiroshima@math.kyushu-u.ac.jp} and
 Noriaki Teranishi\footnote{
Higher Education Support Center, Hokkaido University of Science}}
\date{\today}
\maketitle
\begin{abstract}
A conjugate operator $T$ of one-dimensional harmonic oscillator 
$N$ is defined by 
an operator satisfying canonical commutation relation $[N,T]=-i\one$ 
on some domain but  not necessarily  a dense one. 
Examples of conjugate operators include the angle operator 
$\TA$ and the Galapon operator $\TG$. 
Let $\sT$ denote a set of conjugate operators of $N$ of the form 
$T_{\omega,m}=\frac{i}{m}\log(\omega\one-L^m)$ with $(\omega, m)\in \overline{\DD}\times (\NN\setminus\{0\})$, where $L$ is a shift operator and $\DD$ denotes the open unit disc in the complex plane $\CC$.   
A classification of $\sT$ 
is  given as $\sT=\sT_{\{0\}}\cup\sT_{\DD\setminus\{0\}}\cup
\sT_{\partial \DD}$, where 
$\TA\in\sT_{\{0\}}$ and $\TG\in \sT_{\partial \DD}$. 
The classification is specified by 
a pair of parameters $(\om,m)\in\CC\times\NN$. 
Finally the time evolution 
$T_{\om,m}(t)=e^{itN}
T_{\om,m}e^{-itN}$ 
for 
$T_{\om,m}\in\sT$ 
is investigated, and it is shown that 
$T_{\om,m}(t)$ is periodic with respect to~$t$.
\end{abstract}
\setlength{\baselineskip}{15pt}

\section{Introduction}
\label{pre}
\subsection{Preliminary}
Let us consider conjugate operators $T$ of 1D-harmonic oscillator defined as
\[N=\half (p^2+q^2-\one).\] 
Here
$p=-i\frac{d}{dx}$ is the momentum operator and $q=M_x$ is the multiplication operator by $x$. 
Both are self-adjoint operators on $\LR$. 
$T$  satisfies the canonical commutation relation $[N,T]=-i\one$. 
The commutator $[A,B]$ of linear operators $A$ and $B$ is  defined by 
\begin{equation*}
[A,B]=AB-BA
\end{equation*}
on the domain  
$\mathrm \rD(AB)\cap \mathrm \rD(BA)$. 
Here $\rD(A)$ denotes the domain of $A$. 
In physics, the relationship 
between position and momentum as well as   
between energy and time are often considered to be formally complementary.
Since the quantized position $q$ and momentum $p$ 
satisfy the canonical commutation relation $[q,p]=-i\one$, 
if $N$ is interpreted as the energy of a quantum system, 
$T$ can be viewed as a quantization of time. 
Consequently, 
$T$  is frequently referred to as the time operator associated with $N$. 
However, this paper does not engage in any philosophical discussions regarding the concept of time. 
It is important to emphasize that the term \lq\lq time operator" 
is used solely as a convenient label. 
Precise mathematical definitions of time operators and conjugate operators are provided in Definition~\ref{conj}.

In the previous paper \cite{HT22a} we consider two special conjugate operators of $N$, 
namely the angle operator $\TA$ and the Galapon operator $\TG$ 
from a mathematical point of view. 
While $\TA$ and $\TG$ have been mainly studied so far 
from a physical point of view, e.g., 
\cite{MT45, AB61,SG64,CN68,bau83,LLH96,LLH98,SV98},  
to the best of our knowledge, 
there are no firm or robust results regarding the relationship  
between $\TA$ and $\TG$ in a purely mathematical setting. 
In this paper we comprehensively investigate conjugate operators including 
$\TA$ and $\TG$, 
and 
establish a relationship between them. 
Furthermore we classify conjugate operators using a pair of 
parameters $(\om, m)\in \CC\times\NN$ and define three disjoint classes of conjugate operators. 

\subsection{Angle operators and Galapon operators}
Define the annihilation operator $a$ and the creation operator $a^\ast$ in $\LR$ by 
\begin{align*}
a=\frac{1}{\sqrt 2}(q+ip),\quad \add=\frac{1}{\sqrt 2}(q-ip), 
\end{align*}
respectively. 
The canonical commutation relation (CCR) 
\begin{align}\label{ccr}
[a,\add]=\one,\quad [a,a]=0=[\add,\add]
\end{align} 
holds true on a dense subspace of $\LR$. 
The number operator is defined by 
$\add a $ which is actually the harmonic oscillator 
\begin{align}\label{number}
\add a=N,
\end{align}
and $N$ is self-adjoint on $\rD(N)=\rD(p^2)\cap \rD(q^2)$. 
The normalized ground state of $N$ is given by 
\begin{align}\label{1.2}
v(x)=\pi^{-1/4}e^{-x^2/2}.\end{align} 
Note that $a v=0$ and $N v=0$. 
The normalized eigenvectors $v_n$ are given by 
\begin{align}\label{en}
v_n=\frac{1}{\sqrt{n!}} {\add}^n v\quad n\geq0.
\end{align}
Here we write $a^\ast{}^n$ for $(a^\ast)^n$.
It satisfies that 
$Nv_n=nv_n$ and $\s(N)=\NN$. 
Here $\s(N)$ denotes the spectrum of $N$. 
Note that notation $\NN$ describes 
$\{0,1,2,\ldots,\}$ including zero in this paper.

We recall the definitions of conjugate operators and time operators. 
\begin{definition}[Conjugate operators and time operators]
\label{conj}
If a self-adjoint operator $A$ on a Hilbert space $\cH$ 
admits an operator $B$ satisfying 
the canonical commutation relation:
\begin{align}\label{abc}
[A,B]=-i\one\end{align}
on $D_{A,B}\subset \rD(AB)\cap \rD(BA)$, where  
$D_{A,B}\neq\{0\}$, 
then $B$ is referred to as a {conjugate operator} of $A$, and $D_{A,B}$ 
is called a CCR-domain of \kak{abc}.
Furthermore if $B$ is symmetric, then $B$ is referred to as a {time operator} of $A$. \end{definition}
\begin{remark}
Conjugate operators are not necessarily densely defined. However, time operators are densely defined because a time operator is symmetric.
\end{remark}
We shall construct a conjugate operator $T$ of $N$, i.e., 
\begin{align}
\label{n}
[N, T]=-i \one
\end{align}
on some domain.

We investigate  the operator 
$-\half \left( \arctan\left(q\f p\right) +\arctan\left(pq\f\right)\right)$ 
in this paper. 
Unfortunately, the domains of $\arctan\left(q\f p\right)$ and $\arctan\left(pq\f\right)$
intersect trivially, i.e., 
$$\rD(\arctan\left(q\f p\right))\cap \rD(\arctan\left(pq\f\right))=\{0\}.$$ 
As a result, $-\half \left( \arctan\left(q\f p\right) +\arctan\left(pq\f\right)\right)$ becomes trivial. 
To address this issue  we define a direct sum of unbounded operators as follows. 
\begin{definition}
Let $A$ and $B$ be linear operators on $\mathcal H$. 
Let $\overline{\rD(A)}$ denote the closure of the domain of $A$, 
and similarly, let $\overline{\rD(B)}$ denote the closure of the domain of $B$. 
Suppose that 
$A$ is reduced by $\overline{\rD(A)}$ and 
$B$ is also reduced by $\overline{\rD(B)}$. I.e., 
$A{\rD(A)}\subset \overline{\rD(A)}$ and 
$B{\rD(B)}\subset \overline{\rD(B)}$. 
In addition, 
assume that the Hilbert space $\mathcal{H}$ is decomposed as $\mathcal{H} = \overline{\rD(A)} \oplus \overline{\rD(B)}$. Then, we define 
the densely defined operator $A\oplus B$ on $\mathcal H$ by 
\begin{align*}
	&\rD(A\oplus B)=\rD(A)\oplus\rD(B),\\
	&A\oplus B(\varphi\oplus \psi)=A\varphi\oplus B\psi,\quad \varphi\in\rD(A),\ \psi\in\rD(B).
\end{align*}
\end{definition}
In the previous paper \cite{HT22a} we investigate 
the angle operator e.g., \cite{MME07} given by 
\begin{align}\label{ta1}
\TA=-\half \left( \arctan\left(q\f p\right)\oplus\arctan\left(pq\f\right)\right) \end{align}
 and the so-called Galapon operator \cite{AM08,ara09,gal02a,gal02b} 
 defined by 
 \begin{align}\TG =i\sum_{n=0}^\infty\left( \sum_{m\neq n}\frac{(v_m, \cdot) }{n-m}v_n\right).
 \end{align}
It is known that $\TG$ is a bounded self-adjoint operator, and hence
$\TG$ is a time operator of $N$. 
In contrast,  $\TA$ is 
not symmetric, and hence $\TA$ is not a time operator of $N$.  
Moreover $[N,\TG]=-i\one$ holds on a dense domain, 
whereas $[N, \TA]=-i\one$ on a non-dense domain.

Many of conjugate operators of $N$ discussed in this paper
are not densely defined and hence not symmetric. 
To address this, we extend the usual canonical commutation relation \kak{n} to 
the so-called ultra-weak canonical commutation relation.
See Definition \ref{def-uwt}. 
For the sesqui-linear form $\cT$, 
the ultra-weak canonical commutation relation is defined as
\begin{align}\label{kt}
\cT [N\psi,\varphi] -\ov{\cT [N\varphi, \psi]}=-i(\psi,\varphi), 
\end{align}
where $\cT[\psi,\varphi]$ is linear in $\varphi$ and anti-linear in $\psi$. 
Importantly $\cT$ is not required to be densely defined on $\LR\oplus \LR$. 
Using this framework, we can define an ultra-weak time operator 
$\cT_A$ associated with $\TA$. It was shown in \cite{HT22a} that 
$\cT_A$ satisfies \kak{kt} on a dense domain.
See \kak{T6A}.

\subsection{Significance of research on time operators}
The study of time operators addresses one of the most fundamental and unresolved questions in quantum theory since 1925. 
Pauli's celebrated argument suggested that a self-adjoint time operator canonically conjugate to a semibounded Hamiltonian cannot exist, apparently excluding time from the operator formalism of quantum mechanics. 
However, modern developments have revealed that this limitation is not absolute: by extending the framework of observables from self-adjoint operators to positive operator-valued measures e.g., \cite{gal02a, gal02b} and to strong time operators e.g., \cite{AM08, miy01}, consistent notions of time observables can indeed be defined. 

From the viewpoint of the uncertainty principle derived from the canonical commutation relation, the existence of a time operator provides a concrete operator-theoretic realization of the time-energy uncertainty relation; see, for example, the Kennard inequality \cite{gia97} below: 
\begin{proposition}[Kennard inequality]
Let $A$ and $B$ be self-adjoint operators on a Hilbert space $\cK$.
Fix a unit vector $\psi\in\mathcal H$ such that $\psi\in \rD(AB)\cap \rD(BA)$
and $\psi\in\rD(A)\cap\rD(B)$, and set
$\langle X\rangle_\psi = (\psi, X\psi)$, 
$\tilde A = A-\langle A\rangle_\psi \one$ and 
$\tilde B = B-\langle B\rangle_\psi \one$. 
Define the standard deviations
$\sigma_A(\psi)=\|\tilde A\psi\|$ and $\sigma_B(\psi)=\|\tilde B\psi\|$.
Then
\[
\sigma_A(\psi)\sigma_B(\psi)\;\ge\;\frac{1}{2} \big|\langle [A,B]\rangle_\psi \big|.
\]
\end{proposition}

Moreover, a general uncertainty principle is introduced in \cite{AH17}. 
Consequently, research on time operators not only deepens the mathematical understanding of the uncertainty principle and the operator structure of quantum theory, but also offers an operational bridge between abstract formalism and measurable temporal phenomena.

\subsection{Outline of the paper}
In this paper, for notational simplicity, we use symbol $\eln$ to denote the space $\ell^2(\NN)$,  
which consists of  square-summable complex-valued  sequences on $\NN$. 
We also identify $\LR$ with $\eln$
and 
consider conjugate operators of $N$ not on $\LR$ but on $\eln$. 
The key ingredient of our analysis involves  
the shift operators $L$ and $L^\ast$ on $\eln$. 
Here $L$ represents  the left shift, 
and $L^\ast$, being  the adjoint of $L$,  represents the right shift on $\eln$. 
The main part of this paper is presented in Sections \ref{3}-\ref{6}, where 
we investigate (1)-(5) below. \\
(1) We show that $\TA$ and $\TG$ can be represented in terms of 
$L$, $L^\ast$ and $N$ as 
\begin{align*}
&\TG=i\left\{ \log(\one-L)+\log(\one-L^\ast)\right\},\\
&\TA=\frac{i}{2}\left\{
\log\left( \sqrt{\frac{N+2\one}{N+\one}}L^2\right)\oplus
\log\left( \sqrt{\frac{ N+\one}{{N+2\one}}}L^2\right)
\right\}
\end{align*}
in Theorems \ref{ang} and \ref{UT}.\\
(2) We generalize $\TG$ and $\TA$ as 
\begin{align*}
&i\left\{ \log(\one-g_NL)+\log\left(\one-L^\ast g_N\f\right)\right\},\\
&\frac{i}{2}\left\{
\log\left( g_{N+2}L^2\right)\oplus
\log\left( \tilde g_{N+2}L^2\right)
\right\},
\end{align*}
respectively, in Sections \ref{3} and \ref{4}. 
Here $g_.$ is a map from $\NN$ to $\CC$. \\
(3) 
We construct a general class of conjugate operators in Section \ref{5}. 
Let
\begin{align*}L_{\om,m}=\om\one -{L}^m,\quad (\om,m)\in\CC\times \NN.\end{align*}
We define 
$T_{\om,m}$ by 
\begin{align}
\label{gg}
T_{\om,m}=\frac{i}{m}\log L_{\om,m}.
\end{align}
We can verify that $T_{\om,m}$ are conjugate operators of $N$. 
Let $\DD$ denote  the open unit disc in the complex plane $\CC$. 
For $\om\in\partial\DD$ it can be shown that $T_{\om,m}$ are bounded and admit  
dense CCR-domains. We refere to  
\begin{align*}T_{\om,m},\quad m\geq1, \ \om\in\partial\DD\end{align*}
 as general Galapon operators. In particular it can be seen 
 that $\TG=T_{1,1}+T_{1,1}^\ast$. 
On the other hand 
general angle operators are introduced as
\begin{align}
\label{ga}
T_{\om,m},\quad m\geq1,\ \om=0. 
\end{align}
Let
\begin{align*}\sT=\{T_{\om,m}\mid \om\in\overline{\DD}, m\geq1\}.\end{align*}
We divide $\sT$ into three disjoint families based on $\om\in\overline{\DD}$. 
We have 
\begin{align*}\sT=\sT_{\{0\}}\cup\sT_{\DD\setminus\{0\}}\cup\sT_{\partial \DD}.\end{align*}
Here 
\begin{align*}
&\sT_{\{0\}}=\{T_{\om,m}\mid \om=0,m\geq1\},\\
&\sT_{\DD\setminus\{0\}}=\{T_{\om,m}\mid 0< |\om|<1,m\geq1\},\\
&\sT_{\partial \DD}=\{T_{\om,m}\mid |\om|=1,m\geq1\}.
\end{align*}
We study CCR-domains for each conjugate operator in $\sT$. \\
(5)
Let $H$ be a self-adjoint operator. 
Then the strong time operator $T_H$ is defined by 
the weak Weyl relation
$$T_He^{-itH}\supset e^{-itH}(T_H+t),\quad t\in\RR.$$
It is established that 
if $T_H$ is a strong time operator for $H$, then 
the spectrum of $H$ is 
purely continuous. 
We can show  in Section \ref{6} that conjugate operators in $\sT$ satisfy 
a discrete version of the weak Weyl relation. 
Furthermore we show that the time evolution 
$T_{\om,m}(t)=e^{itN} T_{\om,m}e^{-itN}$ is periodic in $t$ with period $2\pi/m$.

\section{Technical tools}
\subsection{\textbf{Super coherent vectors}}
We define the exponential operator $e^A$ for a non self-adjoint operator $A$ in a Hilbert space $\cH$ by the geometric series:
\begin{align*}
&\rD\left( e^A\right)=\left\{f\in \bigcap_{k=0}^\infty \rD(A^k)\ \middle|\ \lim_{K\to\infty}\sum_{k=0}^K\frac{1}{k!}A^kf\ \text{exists}\right\},\\
&e^Af=\sum_{k=0}^\infty \frac{1}{k!}A^kf,\quad f\in\mathrm D\left(e^A\right).
\end{align*}
It should be written 
as $\sum_{k\in\mathbb N} \frac{1}{k!}A^k$ for $e^A$, but we write $e^A$ for the notational simplicity. 
We define the set $\cC$ consisting of coherent vectors by
\begin{align*}\cC=\mathrm{LH} \left\{e^{\beta \add} v\ \Big|\ \beta\in\CC\right\},\end{align*}
where $v$ is given by \kak{1.2} and $e^{\beta \add} $ is an unbounded operator. Note that 
$v\in \rD(e^{\beta \add})$ for any $\beta\in\CC$, 
and it is given by 
\begin{align*}e^{\beta \add}v(x)=\pi^{-1/4}e^{\beta^2/2}\exp\left( -\frac{(x-\sqrt 2\beta)^2}{2}\right) .\end{align*}
It is well known that $\cC$ is dense in $\LR$. 
We define $w_\beta=e^{-|\beta|^2/2}e^{\beta \add} v$, where $e^{-|\beta|^2/2}$ 
is the normalization constant such that $\|w_\beta\|=1$. 
It is shown that 
$w_\beta\in \rD(\add ^n)$ and ${\add}^n w_\beta=e^{-|\beta|^2/2} e^{\beta \add} {\add}^n v$ for all $n\in\NN$. 
Moreover we obtain that
\begin{align*}aw_\beta=\beta w_\beta.\end{align*}
Thus $w_\beta$ is an eigenvector of $a$ corresponding to eigenvalue $\beta\in\CC$. 
We also see that 
$e^{\gamma a}w_\beta=e^{\gamma\beta}w_\beta$ for any $\beta, \gamma\in\CC$.

Next let us consider vectors of the form 
$e^{\beta {\add}^2/2}v$. 
Exponent of ${\add}^2$ is also unbounded. 
While it is clear that 
$e^{\beta \add} v\in\LR$ for any $\beta\in\CC$, 
it is not immediately obvious that $e^{\beta {\add}^2/2}v\in \LR$. 
\begin{lemma}
\label{11}
Let $\beta\in\CC$. 
Then 
$v\in\rD\left( e^{\beta \add ^2/2}\right) $ if and only if $|\beta|<1$. Moreover, $e^{\beta \add^2/2}v\in\rD(\add^n)$ for any $|\beta|<1$ and $n\in\NN$.
\end{lemma}
\proof
Let $|\beta|<1$. 
By the Rodrigues formula and orthogonality of the Hermite polynomials $H_k$,
\begin{align*}
\left\|\sum_{k=0}^K\frac{1}{k!}\left( \frac{\beta}{2}\right) ^k{\add}^{2k}
v\right\|^2
=\left\|\sum_{k=0}^K\frac{\beta^k}{2^{2k} k!}H_{2k}v\right\|^2
=\sum_{k=0}^K\frac{|\beta|^{2k}}{(2^{2k} k!)^2}2^{2k}(2k)!
=\sum_{k=0}^K\frac{(2k-1)!!}{(2k)!!}|\beta|^{2k}.
\end{align*}
Let $K\to\infty$.
Then 
$\|e^{\beta \add^2/2}v\|^2=(1-|\beta|^2)^{-1/2}$
holds. 
This shows that $v\in \rD\left( e^{\beta \add ^2/2}\right) $ if and only if $|\beta|<1$.
The statement that $e^{\beta\add^2/2}v\in \rD\left( \add^n \right) $ is similarly proven.
\qed
We call $e^{\beta {\add}^2/2}v$ a super coherent vector 
and denote the linear hull of super coherent vectors by 
\begin{align*}\cS=\mathrm{LH}\left\{e^{\beta {\add}^2/2}v\ \Big|\ |\beta|<1\right\}.\end{align*}
The vectors contained in $\cS$ play an important role in studying 
the angle operator. 

\subsection{Wiener-It\^o decomposition}
As mentioned in Section \ref{pre} 
we shall investigate conjugate operators of $N$ on $\eln$ for $\LR$, which 
is called the particle-number representation. 
The first task is to clearly establish the identification between $\eln$ and $\LR$, 
and to transform the operators $a$ and $a^\ast$ on $\LR$ into operators on $\eln$. 
By the Wiener-It\^o decomposition we have 
\begin{align*}
\LR=\bigoplus_{n=0}^\infty L_n,
\end{align*}
where $L_n=\mathrm{LH}\{v_n\}$ 
is the one-dimensional linear subspace spanned by eigenvectors $v_n$ of $N$ corresponding to  eigenvalue $n\in\NN$. 
Let $P_n$ be the projection onto $L_n$. Then $N=\bigoplus_{n=0}^\infty nP_n$ is the spectral decomposition of $N$. 
For each $f\in \LR$, there exists $(c_n)_{n\in\NN}\in\eln $ 
so that 
$f=\sum_{n=0}^\infty c_n v_n$. 
In fact $c_n=(v_n,f)$. 
We often write $(c_0,c_1,c_2,\ldots)$ for $(c_n)_{n\in\NN}$. 
By the unitary map 
\begin{equation}\label{unitary}
U f=(c_n)_{n\in\NN},
\end{equation}
we can identify $\LR$ with $\eln $. 
We set $\xi_n=Uv_n\in\eln $. 
It is seen that 
\begin{align*}
&UaU^\ast\colon \xi_n\mapsto \sqrt n\xi_{n-1},\quad n\geq1,\\
&U\add U^\ast\colon \xi_n\mapsto \sqrt{n+1}\xi_{n+1},\quad n\geq0.
\end{align*}
For the notational convenience, 
we write 
$a$ for $UaU^\ast$. 
Furthermore 
$Uv$ is denoted by $\Omega$, where 
$\Omega=(1,0,0,\ldots)$ and 
$\xi_n=(0,\ldots,0,\stackrel{n+1_{th}}{1},0,\ldots)$ for $n\geq0$. 
Note that $\xi_0=\Omega$. 
In the following, our investigation is transformed onto $\eln $ instead of 
on $\LR$ and 
we fix $\{\xi_n\}_{n\in\NN}$ as a complete orthonormal system of $\eln $.
Let $\ell_\mathrm{fin}^2$ be the finite particle subspace defined by 
\begin{align*}\ell_\mathrm{fin}^2=\left\{\sum_{n=0}^m c_n\xi_n
\ \middle|\ m\in\NN,\ c_n\in\CC \right\}.\end{align*}
This subspace is dense, 
and 
commutator $[a,\add]$ and 
polynomials in $a$ and $a^\ast$ are well defined on this space. 
Thus algebraic computations involving $a$ and $a^\ast$ can be performed  on 
$\ell_\mathrm{fin}^2$. 

\subsection{Shift operators $\bm{L}$ and $\bm{L^\ast}$ on $\bm{\eln}$}
\label{LL}
Let $L$ be the left-shift and its adjoint $L^\ast$ the right-shift on $\eln $, 
which are defined by 
\begin{align*}
& L\xi_n=\begin{cases} \xi_{n-1} & n\geq1,\\ 0 & n=0,\end{cases}\\
& L^\ast \xi_n=\xi_{n+1}.
\end{align*}
We can observe the following relations: 
\begin{align*}
&LL^\ast=\one,\\
&L^\ast L=\one-P_{\{\Omega\}},
\end{align*} 
where $P_{\{\Omega\}}$ is the projection onto the one-dimensional subspace spanned by $\Omega$. 
In general any closed operator $A$ can be decomposed 
as $A=V|A|$, where $|A|=(A^\ast A)^{1/2}$ is a positive self-adjoint operator
 and 
$V$ is a partial isometry such that the initial space is $(\ker A)^\perp$ 
and the final space $\ov{\ran A}$. 
This is called the polar decomposition of $A$. 
The operator $V$ is uniquely determined by these properties, along with  
the condition $\ker |A|=\ker V$.
For the operators $L$, $L^\ast$ and $N$, 
both $a$ and $a^\ast$ can be represented as 
\begin{align*}
&a=L\sqrt N=\sqrt{N+\one}L,\\
&\add=L^\ast \sqrt{N+\one}=\sqrt N L^\ast.
\end{align*} 
These are just the polar decompositions of 
the closed operators $a$ and $\add$. 
Note that $N=\add a$ and $N+\one=a\add$. 
We also observe that
\begin{align*}
&\ker a=\{c\Omega\mid c\in\CC \}=\ker \sqrt N,\\
&\ker \add=\{0\}=\ker \sqrt{N+\one}.
\end{align*} 
$L$ is a partial isometry such that the initial space is 
$(\ker a)^\perp =\{c\Omega\mid c\in\CC \}^\perp\cong\bigoplus_{n=1}^\infty L_n$ and the final space 
$\eln $. 
We also see that 
\begin{align}
[N,L^\ast]\subset L^\ast,\quad [N, L]\subset -L.
\end{align}
Thus
$[N, {L^\ast}^k]\subset k{L^\ast}^k$ and $[N, L^k]\subset -kL^k$ hold true. 
Let 
$\DD=\{z\in\CC\mid |z|<1\}$
 be the open unit disc in $\CC$. 
We observe that 
$$e^{\alpha NL^\ast}\Omega=(1,\alpha,\alpha^2,\alpha^3,\ldots),\quad
\alpha\in \DD.$$
It follows from this that 
\begin{align}\label{al}
Le^{\alpha NL^\ast}\Omega=\alpha e^{\alpha NL^\ast}\Omega,\quad
\alpha\in \DD.
\end{align}
Thus $e^{\alpha NL^\ast}\Omega$ is an eigenvector of $L$ corresponding to the eigenvalue $\alpha$. 
\begin{lemma}\label{el}
We have 
$\s(L)=\s(L^\ast)=\ov{\DD}$, 
$\s_{\rm p}(L)=\DD$ and $\s_{\rm p}(L^\ast)=\emptyset$. 
\end{lemma}
\proof
Since $\|L\|\leq 1$, 
$\s(L)\cup\s(L^\ast)\subset \ov{\DD}$. 
By \kak{al}, we see that $\DD\subset \s_{\rm p}(L)$ and $\s(L)=\s(L^\ast)=\ov{\DD}$. 
Let $|\alpha|=1$. 
The relation $L\varphi=\alpha \varphi$ implies $\varphi=(\alpha^n)_{n\in\NN}$ but $\varphi\not\in \eln$. 
Hence 
$\s_{\rm p}(L)= \DD$. 
Let $|\alpha|\leq1$ and $L^\ast \varphi=\alpha \varphi$. 
Then $(L^\ast \varphi)_0=0=\alpha \varphi_0$ and hence 
$(L^\ast \varphi)_1=0$. Repeating this procedure, we see that $\varphi=0$. Thus we have
$\s_{\rm p}(L^\ast)=\emptyset$.
\qed
Let $k\in \NN$. 
Similarly to the proof of Lemma \ref{el} 
we can see that, for any $k\alpha\in \DD$, $e^{\alpha N {L^\ast}^k}\Omega$ 
is an eigenvector of $L^k$. 
Since 
\begin{align*}e^{\alpha N {L^\ast}^k}\Omega=(1,
\underbrace{0,\ldots,0}_{k-1},k\alpha,\underbrace{0,\ldots,0}_{k-1},(k\alpha)^2,\underbrace{0,\ldots,0}_{k-1},(k\alpha)^3,\ldots),\end{align*} 
it is given by 
\begin{align}
L^ke^{\alpha N {L^\ast}^k}\Omega=k\alpha e^{\alpha N {L^\ast}^k}\Omega.
\end{align}
We can also extend this to a general version. 
Let $f$ be a complex-valued function on $\NN$, and set 
 $f(N)=f_N$. 
Then 
\begin{align}
\label{teranishi}
e^{\alpha f_N {L^\ast}^k}\Omega=\biggl(1,\underbrace{0,\ldots,0}_{k-1},f(k)\alpha,\underbrace{0,\ldots,0}_{k-1},\frac{f(2k)f(k)}{2!}\alpha^2,\underbrace{0,\ldots,0}_{k-1},\frac{f(3k)f(2k)f(k)}{3!}\alpha^3,\ldots\biggr).
\end{align}
In Sections 
 \ref{3} and \ref{4}, for a given function $f$, we shall find a function $g$ such that 
 $e^{\alpha f_N {L^\ast}^k}\Omega$ is an eigenvector of $g_{N+k}L^k$.

\section{\textbf{Phase operators}}
We give the definition of $\log A$ for a linear operator $A$. 
It is emphasized that 
$A$
 is not necessarily self-adjoint.
\begin{definition}[$\log A$]\label{def}
Let $A$ be a linear operator on a Hilbert space $\cH$. 
We define $\log A$ by 
\begin{align*}
&\rD(\log A)=\left\{f\in \bigcap_{k=0}^\infty\rD(A^k)\ \middle|\ 
\lim_{K\to\infty}\sum_{k=1}^K \frac{1}{k}(\one-A)^k f\text{ exists } \right\},\\
&\log Af=-\sum_{k=1}^\infty \frac{1}{k}(\one-A)^kf,\quad f\in\rD(\log A).
\end{align*}
\end{definition}
Note that in general 
\begin{align}
&\label{log1}
\log (AB)\neq \log A+\log B,\\
&\label{log2}
\log\alpha A\neq \log \alpha +\log A,\quad \alpha\in\CC.
\end{align}
In the physics the so-called phase operator  $\rho$ is formally given by $a=e^{i\rho}\sqrt N$. 
Therefore we define the phase operator $\rho$ on $\ell^2$ by 
\begin{align}\label{phase}
\rho=-\frac{i}{2}(\log a-\log\add). 
\end{align}
The phase operator $\rho$ has been studied in numerous literatures, e.g., 
\cite{SG64, LLH96, PG89} as the conjugate operator of $N$, 
but it does not appear to be obvious in the definition of $\log a$ and $\log a^\ast$. 
\begin{lemma}\label{ng}
We have 
$\rD(\log a)\cap\ell_\mathrm{fin}^2=\{0\}$ and 
$\rD(\log \add)=\{0\}$. In particular, $\rD(\rho)=\{0\}$. 
\end{lemma}
\proof
Let $\psi=\sum_{l=0}^mc_l\xi_l\in \ell_\mathrm{fin}^2$ with $c_m\neq0$. 
Note that 
$({a^\ast}^k\xi_m,\psi)=0$ for $k\geq1$. 
Since 
\begin{align*}
\lim_{K\to\infty}\left|\left(\xi_m, \sum_{k=1}^K\frac{1}{k}(\one-a)^k\psi\right)\right|
=\lim_{K\to\infty}\left|\sum_{k=1}^K\frac{1}{k}\left((\one-a^\ast)^k\xi_m,\psi\right)\right|
=\lim_{K\to\infty}\left|(\xi_m, \psi)\right|\sum_{k=1}^K\frac{1}{k}=\infty,
\end{align*}
we see that $\rD(\log a)\cap\ell_\mathrm{fin}^2=\{0\}$.
Let $\psi'=\sum_{l=m'}^\infty c_l\xi_l\in \ell^2$ with $c_{m'}\neq0$. 
Similarly, it follows that 
\begin{align*}
\lim_{K\to\infty}\left|\left(\xi_{m'}, \sum_{k=1}^K\frac{1}{k}(\one-a^\ast)^k\psi'\right)\right|=
\lim_{K\to\infty}\left|(\xi_{m'}, \psi')\right|\sum_{k=1}^K\frac{1}{k}=\infty.
\end{align*}
Hence $\rD(\log \add)=\{0\}$.
\qed

From Lemma \ref{ng} we can see that 
$\rho$ can not be defined on $\ell_\mathrm{fin}^2$. 
This fact poses a significant challenge when considering the phase operator 
$\rho$ on $\eln$. 
Therefore, when investigating $\rho$, 
it is crucial to carefully consider its domain. 

\section{Angle operators}
\label{3}
\subsection{\textbf{Angle operators by $\bm{a}$ and $\bm{a^\ast}$}}
It is stated that a relationship between the angle operator $\TA$ and the phase operator $\rho$ is given by $\TA=(\pi/2)\one-\rho+G(N)$ in 
e.g., \cite[(32)]{LLH96}. Here $G(N)$ is an appropriate function of $N$. 
However it seems to be formal and as observed in Lemma \ref{ng} the phase operator $\rho$ is not well defined on $\ell_\mathrm{fin}^2$. 

To establish a rigorous relationship between 
phase operator $\rho$ and the angle operator $\TA$, 
we express $\TA$ in terms of the creation and annihilation operators. 
Define
\begin{align*}\phi=\frac{1}{\sqrt2} (a+\add).\end{align*}
We also define two disjoint subspaces of $\LR$ by 
\begin{align*}
&\cC_0=\mathrm{LH}\left\{e^{-\alpha x^2/2}\ \Big|\ \alpha\in(0,1)\right\},\\
&\cC_1=\mathrm{LH}\left\{x e^{-\alpha x^2/2}\ \Big|\ \alpha\in(0,1)\right\}.
\end{align*}
Note that $\cC_0\perp \cC_1$ and 
$\cC_0+\cC_1$ is dense. 
 In this paper, $\#$ denotes $0 \mbox { or } 1$. 
We transform $\cC_\#$ to the subspaces of $\eln$ using the unitary map $U$ defined in 
\eqref{unitary}. 

\begin{lemma}\label{unitary1}
Suppose that $0< \alpha<1$. Then
\begin{align*}
&Ue^{-\alpha x^2/2}=\pi^{1/4}\sqrt{\frac{2}{1+\alpha}}\exp\left( \frac{1-\alpha}{1+\alpha}\frac{\add ^2}{2}\right) \Omega,\\
&Uxe^{-\alpha x^2/2}=\pi^{1/4}\sqrt{\frac{2}{1+\alpha}}{\frac{\sqrt 2}{1+\alpha}}\add \exp\left( \frac{1-\alpha}{1+\alpha}\frac{\add ^2}{2}\right) \Omega.
\end{align*}
\end{lemma}
\proof
We set $\beta=(1-\alpha)/2$.
Since $UxU^\ast=\phi$ and $Ue^{-x^2/2}=\pi^{1/4}\Omega$, 
we have 
\begin{align*}
Ue^{-\alpha x^2/2}
&=Ue^{-(\alpha-1) x^2/2}U^\ast Ue^{-x^2/2}
=\pi^{1/4} \sum_{n=0}^\infty\frac{1}{n!}\beta^n\phi^{2n}\Omega\\
&=\pi^{1/4} \sum_{k=0}^\infty\left( \sum_{n\geq k}\frac{1}{(n-k)!}\frac{(2n-1)!!}{2^{n-k}(2k-1)!!}\beta^n\right) \frac{1}{k!}\frac{\add ^{2k}}{2^k}\Omega\\
&=\pi^{1/4} \sum_{k=0}^\infty\frac{1}{\sqrt{1-\beta}}\left( \frac{\beta}{1-\beta}\right) ^k\frac{1}{k!}\frac{\add ^{2k}}{2^k}\Omega\\
&=\pi^{1/4}\sqrt{\frac{2}{1+\alpha}}\exp\left( \frac{1-\alpha}{1+\alpha} \frac{\add ^2}{2}\right) \Omega.
\end{align*}
Hence we have 
\begin{align*}
U
xe^{-\alpha x^2/2}=
\pi^{1/4}\sqrt{\frac{2}{1+\alpha}}\phi\exp\left( \frac{1-\alpha}{1+\alpha} \frac{\add ^2}{2}\right) \Omega
=\pi^{1/4}\sqrt{\frac{2}{1+\alpha}}\frac{\sqrt 2}{1+\alpha}\add \exp\left( \frac{1-\alpha}{1+\alpha} \frac{\add ^2}{2}\right) \Omega. 
\end{align*}
Then the lemma is proven.
\qed

Disjoint sets of super coherent vectors are defined by
\begin{align*}
&\cS_0=\mathrm{LH} \left\{e^{\beta \add ^2/2} \Omega\ \Big|\ \beta\in (0,1)\right\},\\
&\cS_1=\mathrm{LH} \left\{ \add e^{\beta \add ^2/2} \Omega\ \Big|\ \beta\in (0,1)\right\}.
\end{align*}
Note that 
$\cS_0\perp \cS_1$ and 
$\cS_0\cup \cS_1$ is dense. 
Lemma \ref{unitary1} also shows that $U\cC_\#=\cS_\#$.
In order to express $\TA$ in terms of the operators $a$ and $a^\ast$, 
we need the inverse of the creation operator $\add$, which  is defined by 
\begin{align*}
&\rD\left( \add \f\right) =\left\{\sum_{n=0}^\infty c_n\xi_n\in\eln \ \bigg|\ 
c_0=0\right\}, \\
&\add \f \sum_{n=1}^\infty c_n\xi_n=\sum_{n=1}^\infty \frac{c_n}{\sqrt n}\xi_{n-1}.
\end{align*} 
We consider operators $\log (\add \f a)$ and $\log (a \add \f )$. 
Note that the operator $\add \f$ is well defined on $\ran L^\ast$. 
\begin{lemma}\label{super}
We have (1) and (2). 
\bi
\item [{\rm(1)}]
$\cS_0
\subset \rD\left( \log (\add \f a)\right) $ and 
\begin{align*}\log \left( \add \f a\right) e^{\beta \add ^2/2}\Omega=(\log\beta)e^{\beta \add ^2/2}\Omega,\quad 0<\beta<1.\end{align*}
\item [{\rm(2)}]
$\cS_1\subset \rD\left( \log (a \add \f )\right) $ and 
\begin{align*}\log \left( a\add \f \right) \add e^{\beta \add ^2/2} \Omega=(\log\beta) \add e^{\beta \add ^2/2}\Omega,\quad 0<\beta<1\end{align*}
\ei
\end{lemma}
\proof
Since $e^{\beta \add ^2/2}\Omega$ is an eigenvector of $\add{}^{-1}a$ 
corresponding to the eigenvalue $\beta$, for all $n\in\NN$, we see that 
$(\one-{\add}^{-1}a)^n e^{\beta \add ^2/2}\Omega
=(1-\beta)^n e^{\beta \add ^2/2}\Omega$. 
Hence by the definition of $\log \left( \add \f a\right) $ we obtain that 
$\log \left( \add \f a\right) e^{\beta \add ^2/2}\Omega
=(\log \beta) e^{\beta \add ^2/2}\Omega$. 
The proof of (2) is the same as that of (1). 
\qed

\begin{lemma}
Both 
$\log (\add \f a)$ and $\log (a\add \f)$ are unbounded. 
\end{lemma}
\proof
By Lemma \ref{super}, we can see that 
$e^{\beta \add ^2/2}\Omega$ is an eigenvector of 
$\log \left( \add \f a\right) $ corresponding to eigenvalues 
$\log\beta$.
Similarly 
$\add e^{\beta \add ^2/2}\Omega$ is an eigenvector of 
$\log \left( a\add \f \right) $ corresponding to eigenvalues 
$\log\beta$.
Since $0<\beta <1$, we have
\begin{align*}
\s\left( \log \left(\add \f a\right)\right) \cap \s\left( \log \left(a\add \f\right)\right) \supset(-\infty, 0).
\end{align*}
Hence the lemma follows. 
\qed

We recall the  definition of $\arctan(A)$ for a linear operator $A$.
\begin{definition}
Let $A$ be a linear operator on a Hilbert space $\mathcal H$. 
We define the  linear operator $\arctan(A)$ as follows:
\begin{align*}
&\rD(\arctan(A))=\left\{\varphi\in \bigcap_{k=0}^\infty \rD \left(A^{2k+1}\right)
\ \middle|\ 
\lim_{K\to\infty} 
\sum_{k=0}^K \frac{(-1)^k}{2k+1}A^{2k+1}\varphi\mbox{ exists}\right\},\\
&\arctan(A)\varphi=-\sum_{k=0}^\infty \frac{(-1)^k}{2k+1}A^{2k+1}\varphi,\quad \varphi\in\rD(\arctan(A)).
\end{align*}
\end{definition}
Now we transform $\TA$ to an operator on $\eln$. 
We define the set of even functions in $\LR$ as $L^2_\mathrm{e}(\RR)$
and the set  of odd functions in $\LR$ as $L^2_\mathrm{o}(\RR)$:  
\begin{align*}
&L^2_\mathrm{e}(\mathbb R)=\{ f\in L^2(\mathbb R)\mid f(x)=f(-x) \text{ for all  }x\in \RR\},\\
&L^2_\mathrm{o}(\mathbb R)=\{ f\in L^2(\mathbb R)\mid f(x)=-f(-x) \text{ for all }x\in \RR\}. 
\end{align*}
We define 
\begin{align*}
&\cM_0=\mathrm{LH}\left\{x^{2n}e^{-\alpha x^2/2}\in\LR\ \Big|\ n\in\NN,\ \alpha\in(0,1)\right\},\\
&\cM_1=\mathrm{LH}\left\{x^{2n+1}e^{-\alpha x^2/2}\in\LR\ \Big|\ n\in\NN,\ \alpha\in(0,1)\right\}.
\end{align*}
Note that $\overline{\cM_0}=L^2_\mathrm{e}(\RR)$ and that $\overline{\cM_1}=L^2_\mathrm{o}(\RR)$.
Subspaces $\cM_\#$ are useful subspaces for considering the commutation relation between the harmonic oscillator and the angle operator:
\begin{align*}
&\left[N,-\arctan\left( q^{-1}p\right) \right]=-i\one\quad\textrm{ on }\cM_0,\\
&\left[N,-\arctan\left( pq^{-1}\right) \right]=-i\one\quad\textrm{ on }\cM_1.
\end{align*}
We also define 
\begin{align*}
&\ell^2_\mathrm{e}=\left\{\varphi=\sum_{n\in\mathbb{N}}c_n\xi_n\in\ell^2\ \middle|\ c_{2n+1}=0 \text{ for all }n\in\mathbb N\right\},\\
&\ell^2_\mathrm{o}=\left\{\varphi=\sum_{n\in\mathbb{N}}c_n\xi_n\in\ell^2\ \middle|\ c_{2n}=0 \text{ for all }n\in\mathbb N\right\}
\end{align*}
and 
\begin{align}
\label{NN1}&\cN_0=\mathrm{LH}\left\{a^\ast{}^{2n}e^{\beta a^\ast{}^2/2}\Omega\in\eln\ \Big|\ n\in\NN,\ \beta\in(0,1)\right\},\\
\label{NN2}&\cN_1=\mathrm{LH}\left\{a^\ast{}^{2n+1}e^{\beta a^\ast{}^2/2}\Omega\in\eln\ \Big|\ n\in\NN,\ \beta\in(0,1)\right\}.
\end{align}
It is immediate to see that $UL^2_\mathrm{e}(\mathbb R)=\ell^2_\mathrm{e}$, $UL^2_\mathrm{o}(\mathbb R)=\ell^2_\mathrm{o}$ and $U\cM_\#=\cN_\#$. 
Note that $\cS_0\subset\cN_0\subset \ell^2_\mathrm{e}$ and $\cS_1\subset\cN_1\subset \ell^2_\mathrm{o}$. 
We show algebraic relations we used often in this paper. 
Let $X$ and $Y$ be linear operators. 
We define the map $\ad_X$ by 
$\ad_X(Y)=[X,Y]$. 
Let $A$ and $B$ be linear operators. 
Then the algebraic relation 
\begin{align*}ABf=BAf-\ad_B(A)f=(B-\ad_B)Af\end{align*} 
holds for $f\in\rD(AB)\cap\rD(BA)$. Hence 
$AB^2=(B-\ad_B)^2A$ on $\rD(AB^2)\cap\rD(BAB)\cap\rD(B^2A)$ and then for any $n\in\NN$ and any $f\in\bigcap_{k=0}^n\rD(B^kAB^{n-k})$, 
\begin{align}
\label{N}
AB^nf=(B-\ad_B)^nAf.
\end{align}

\begin{lemma}\label{uag}
The following relations hold: 
\begin{align*}
U\arctan\left( q^{-1}p\right) U^\ast&=-\frac{i}{2}\log\left( \add ^{-1}a\right) \hspace{5pt}\mbox{ on } \cN_0,\\
U\arctan\left( pq^{-1}\right) U^\ast&=
-\frac{i}{2}\log\left( a \add ^{-1}\right) \hspace{5pt}\mbox{ on } \cN_1.\end{align*}
\end{lemma}
\proof
We see that by \cite[Lemma 3.6]{HT22a}
\begin{align*}\arctan\left( q^{-1}p\right) x^{2n}e^{-\alpha x^2/2}
=\frac{i}{2}\left\{\left( x^2-2\frac{d}{d\alpha}\right) ^n\log\left( \frac{1+\alpha}{1-\alpha}\right) \right\}e^{-\alpha x^2/2}.\end{align*}
Then, 
by Lemma \ref{unitary1}, 
\begin{align}
&U\arctan\left( q^{-1}p\right) x^{2n}e^{-\alpha x^2/2}
\nonumber \\
&\label{U}
=-\frac{i}{2}\left\{\left( \phi^2-2\frac{d}{d\alpha}\right) ^n\log\left( \frac{1-\alpha}{1+\alpha}\right) \right\}\pi^{1/4}\sqrt{\frac{2}{1+\alpha}}
\exp\left( \frac{1-\alpha}{1+\alpha}\frac{\add ^2}{2}\right) \Omega.
\end{align}
Let $\bbb =a^\ast{}^{-1}a$. 
Since 
$
\ad_{\phi^2}(\bbb )=-(1+\bbb )^2$ 
on $\cN_0$, we have for any $m\in\mathbb N$ and any analytic function $F$, 
\begin{align*}
\ad_{\phi^2}^m \left( F(\bbb ) \right)=\left\{\left( -(1+x)^2\frac{d}{dx}\right)^m F(x)\right\}\bigg|_{x=\bbb }
\end{align*}
on $\cN_0$. Since 
$\bbb \exp\left( \beta \add ^2/2\right)\Omega=
\beta\exp\left( \beta \add ^2/2\right)\Omega$, we see that 
$\exp\left( \beta \add ^2/2\right)\Omega$ is also an eigenvector of 
$\ad_{\phi^2}^m \left( F(\bbb )\right)$ and 
\begin{align*}\left(\ad_{\phi^2}^m F(\bbb )\right)
\exp\left( \frac{\beta}{2} \add ^2\right)\Omega
=\left\{
\left( -(1+\beta)^2\frac{d}{d\beta}\right)^m F(\beta)\right\}
\exp\left( \frac{\beta}{2} \add ^2\right)\Omega.\end{align*}
In particular 
\begin{equation*}
\left(\ad_{\phi^2}^m F(\bbb )\right)
\exp\left( \frac{1-\alpha}{1+\alpha} \frac{\add ^2}{2}\right)\Omega
=\left\{
\left( 2\frac{d}{d\alpha}\right)^m F\left( \frac{1-\alpha}{1+\alpha}\right) \right\}
\exp\left( \frac{1-\alpha}{1+\alpha} \frac{\add ^2}{2}\right)\Omega.
\end{equation*}
Moreover replacing $F(\bbb )$ with $\log\bbb $, it is obtained that 
\begin{align}\label{M}
\left(
\ad_{\phi^2}^m
\log\bbb \right) \exp\left( \frac{1-\alpha}{1+\alpha}\frac{\ \add^2}{2}\right) \Omega=
\left\{\left( 2\frac{d}{d\alpha}\right)^m\log\left( \frac{1-\alpha}{1+\alpha}\right) \right\}\exp\left( \frac{1-\alpha}{1+\alpha}\frac{\ \add^2}{2}\right) \Omega. 
\end{align}
By \kak{U} and \kak{M}, we have 
\begin{align*}
U\arctan\left( q^{-1}p\right) x^{2n}e^{-\alpha x^2/2}
=-\frac{i}{2}\left\{\left( \phi^2-\ad_{\phi^2}\right)^n\log\bbb \right\}\pi^{1/4}\sqrt{\frac{2}{1+\alpha}}\exp\left( \frac{1-\alpha}{1+\alpha}\frac{\add ^2}{2}\right) \Omega.
\end{align*}
From \kak{N}, we can furthermore see that 
\begin{align*}
U\arctan\left( q^{-1}p\right) x^{2n}e^{-\alpha x^2/2}=&-\frac{i}{2}(\log\bbb ) \phi^{2n}\pi^{1/4}\sqrt{\frac{2}{1+\alpha}}\exp\left( \frac{1-\alpha}{1+\alpha}\frac{\add ^2}{2}\right) \Omega\\
=&-\frac{i}{2}(\log\bbb ) U\left( x^{2n}e^{-\alpha x^2/2}\right) .
\end{align*}
Then the first equality is proven. 
The second equality is similarly proven. \qed
Let 
\begin{align*}
&S_0=\frac{i}{2}\log\left( \add ^{-1}a\right),\\
&S_1=\frac{i}{2}\log\left( a \add ^{-1}\right). 
\end{align*}
Since $\rD(S_0)\subset\ell^2_\mathrm{e}$, $\rD(S_1)\subset\ell^2_\mathrm{o}$ and $\ell^2_\mathrm{e}\perp\ell^2_\mathrm{o}$, as a result $\rD(S_0)\cap \rD(S_1)=\{0\}$. 
Hence
$\rD(S_0+S_1)=\{0\}$. 
Then we define the ultra-weak time operator of $S_0+S_1$. 
For the self-consistency we show the definition of 
ultra-weak time operators below. 
\begin{definition}[Ultra-weak time operator \cite{AH17}]
\label{def-uwt}
Let $H$ be a self-adjoint operator on $\cH $ and 
$D_1$ and $D_2$ be non-zero subspaces of $\cH $.
A sesqui-linear form 
\begin{align*}\cT\colon D_1 \times D_2\to\CC, \quad 
D_1\times D_2\ni (\varphi,\psi)\mapsto \cT [\varphi,\psi]\in\CC\end{align*}
with domain $\rD(\cT )=D_1\times D_2$ 
($\cT [\varphi,\psi]$ is antilinear in $\varphi$ and linear in $\psi$)
is called {\it an ultra-weak time operator} of $H$ if
there exist non-zero subspaces $D$ and 
$E$ of $D_1\cap D_2$ such that
(1)--(3) below hold:
\begin{list}{}{}
\item[(1)] $E \subset \rD(H)\cap D$.
\item[(2)] $\ov{\cT [\varphi,\psi]}=\cT [\psi,\varphi]$ for all $\varphi,\psi\in D$. 
\item[(3)] $HE\subset D_1$ and, for all $\psi, \varphi\in E$, 
\begin{align}
\cT [H\varphi,\psi] -\ov{\cT [H\psi, \varphi]}=-i(\varphi,\psi).\label{TH}
\end{align}
\end{list}
We call $E$ an {\it ultra-weak CCR-domain} 
and $D$ a {\it symmetric domain} of $\cT $. 
\end{definition}
Let us define an ultra-weak time operator of $N$ through $S_0$ and $S_1$. We define 
\begin{align*}
&\cT_0[\varphi,\psi]=\half \left\{(S_0\varphi,\psi)+(\varphi, S_0\psi)\right\},\quad \varphi,\psi\in \cN_0,\\
&\cT_1[\varphi,\psi]=\half \left\{(S_1 \varphi,\psi)+(\psi, S_1\varphi)\right\},
\quad \varphi,\psi\in \cN_1
\end{align*}
and 
\begin{align}
\label{T6}
\cT=\cT_0\oplus \cT_1
\end{align}
with symmetric domain 
$\left( \rD(S_0)\times\rD(S_0)\right) \oplus \left( \rD(S_1)\times\rD(S_1)\right) $, 
where  the direct sum  of \eqref{T6} is defined by 
$\cT[[\varphi_1,\psi_1]\oplus [\varphi_2,\psi_2]]=\cT_0[\varphi_1,\psi_1]+\cT_1[\varphi_2,\psi_2]$. 
On the other hand let us define the ultra-weak time operator $\cT_A$ associated with
the angle operator $\TA$ below. 
Let 
$h_0=\arctan\left( q^{-1}p\right)$ and 
$h_1=\arctan\left( pq^{-1}\right)$. 
We define 
\begin{align*}
&\cT_{A0}[\varphi,\psi]=\half \left\{(h_0\varphi,\psi)+(\varphi, h_0\psi)\right\},\quad \varphi,\psi\in \cN_0,\\
&\cT_{A1}[\varphi,\psi]=\half \left\{(h_1 \varphi,\psi)+(\psi, h_1\varphi)\right\},
\quad \varphi,\psi\in \cN_1
\end{align*}
and 
\begin{align}
\label{T6A}
\cT_A=\cT_{A0}\oplus \cT_{A1}
\end{align}
with the symmetric domain 
$\left( \rD(h_0)\times\rD(h_0)\right) \oplus \left( \rD(h_1)\times\rD(h_1)\right) $. 
\begin{theorem}\label{47}
The ultra-weak time operator $\cT_A$ of the harmonic oscillator on $L^2(\mathbb R)$ is unitary equivalent to $\cT$ on $\ell^2$:
\begin{align}
\label{K}
\cT_{A}[\varphi,\psi]=\cT[U\varphi,U\psi].
\end{align}
\end{theorem}
\proof This follows from Lemma \ref{uag}.
\qed
\begin{remark}
\kak{K} in Theorem \ref{47} can be interpreted as 
a rigorous justification of
$$\half\left( 
\arctan\left( q^{-1}p\right)+
\arctan\left( pq^{-1}\right)\right) 
\cong 
\frac{i}{4}\left\{
\log\left( \add ^{-1}a\right)+
\log\left( a \add ^{-1}\right)\right\}. 
$$
In some literature, however, the following expression is presented:
$$\half\left(
\arctan\left( q^{-1}p\right)+
\arctan\left( pq^{-1}\right)\right)
\cong 
\frac{i}{2}\left\{
\log a-
\log \add \right\}. 
$$
However, by Lemma \ref{ng}, the right-hand side cannot be defined as an operator.
Even if treated as a formal argument, it is dangerous to reason based on this equivalence.
\end{remark}

\subsection{\textbf{Angle operator by shift operators}}
Let $L$ be the left shift operator defined in Section \ref{LL}. 
The angle operator $\TA$ can be represented by $L$ and $N$. 
Let us define 
\begin{align*}
&\LA=
\frac{i}{2}\log\left( \sqrt{\frac{N+2\one}{N+\one}}L^2\right),\\
&\LLA=\frac{i}{2}\log\left( \sqrt{\frac{ N+\one}{{N+2\one}}}L^2\right).
\end{align*}
\begin{theorem}\label{ang}
We have
\begin{align}
\label{SS}
&S_0=\LA \quad\mbox{on }\cN_0,\\
\label{SSS}
&S_1=\LLA\quad\mbox{on }\cN_1,
\end{align}
where 
$\cN_0$ and $\cN_1$ are given by \kak{NN1} and \kak{NN2}, respectively. 
In particular
\begin{align}
\label{uag1}
U\arctan\left( q^{-1}p\right) U^\ast&=\LA \hspace{5pt}\mbox{ on } \cN_0,\\
\label{uag2}
U\arctan\left( pq^{-1}\right) U^\ast&=
\LLA \hspace{5pt}\mbox{ on } \cN_1.
\end{align}
\end{theorem}
\proof
We see that 
\[
\add ^{-1}a=\left(\sqrt N L^\ast\right)^{-1}\sqrt{N+\one}L=L\sqrt{\frac{N+\one}{N}}L
=\sqrt{\frac{N+2\one}{N+\one}}L^2
\]
on $\ran (\one-P_{\{\add\Omega\}})$.
Similarly
\[
a\add ^{-1}=L\sqrt N\left(L^\ast \sqrt{N+\one}\right)\f=
L\sqrt {\frac{N}{N+\one}}L=
\sqrt{\frac{ N+\one}{{N+2\one}}}L^2
\]
on $\ran (\one-P_{\{\Omega\}})$.
From this we have \kak{SS} and \kak{SSS}. 
Relations \kak{uag1} and \kak{uag2} follow from Lemma \ref{uag}. 
\qed

\subsection{\textbf{Generalization of angle operator $\bm{\TA}$}}
We generalize $\LA$ and $\LLA$ to the form 
$i\log \left( g(N)L^2\right)$. 
A fundamental idea is to find eigenvectors $\varphi$ of linear operator 
$\ad_N\bigl(\log \left( g(N)L^2\right)\bigr)$ so that 
$\ad_N\bigl(\log \left( g(N)L^2\right)\bigr)\varphi=a\varphi$ with $a\not=0$. 
Then it follows that 
\begin{align*}\left[N, -\frac{i}{a}\log \left( g(N)L^2\right)\right]=-i\one\end{align*} 
on the CCR-domain $\mathrm{LH}\{\mbox{eigenvectors } \varphi\mbox { of } \ad_N\bigl(\log \left( g(N)L^2\right)\bigr)\}$. 
Hereafter we write $g_N$ for $g(N)$.
Let 
$P_{\geq2}$ be the 
projection onto $\ov{\mathrm{LH}\{\xi_n\mid n\geq2\}}$, 
and $P_0$ be the projection onto 
$\ov{\mathrm{LH}\{\xi_{2n}\mid n\in\NN\}}$. 

\begin{lemma}\label{crfg}
Let $f$ and $g$ be complex-valued functions on $\NN$, 
and $\alpha\in\CC$. 
Suppose that there exists a constant $\beta\in\CC$ such that 
\begin{align}\label{fg}
g_{N+2}f_{N+2}-g_{N}f_NP_{\geq2}=
\beta\one\quad \text{on }P_0\ell_\mathrm{fin}^2.
\end{align}
Then, for all $n\in\NN$, 
we see that 
$\ad_{g_{N+2}L^2}\left( f_N{L^\ast}^2\right)^n=n\beta \left( f_N{L^\ast}^2\right)^{n-1}$ on $P_0\ell_\mathrm{fin}^2$.
\end{lemma}
\proof
The relation \eqref{fg} is equivalent to 
$\bigl[g_{N+2}L^2,f_N{L^\ast}^2\bigr]=\beta\one$ on 
$P_0\ell_\mathrm{fin}^2$. 
This implies 
that $\ad_{g_{N+2}L^2}\left( f_N{L^\ast}^2\right)^n=
n\beta \left( f_N{L^\ast}^2\right)^{n-1}$ on $P_0\ell_\mathrm{fin}^2$.
 \qed

\begin{lemma}\label{radc}
Let $f$ be a complex-valued function on $\NN$. Suppose that 
$|f(2n)|>0$ for all natural number $n\geq1$ and there exists a limit (including infinity)
\begin{align*}
M_f=\lim_{n\to\infty}\frac{n}{|f(2n)|}\leq \infty.
\end{align*}
Then, for all $l,m\in\NN$ and $\alpha\in\CC$ such that $|\alpha|<M_f$, 
\begin{align*}
\Omega\in\rD\left( N^l \left(f_NL^\ast{}^2\right)^me^{\alpha f_NL^\ast{}^2}\right).
\end{align*}
\end{lemma}
\proof
Note
that 
\begin{align*}
\left\|N^l \left(f_NL^\ast{}^2\right)^me^{\alpha f_NL^\ast{}^2}\Omega\right\|^2&
=\sum_{n=0}^\infty 
\frac{|\alpha|^{2n}}{(n!)^2}\left\|
N^l\left( f_NL^\ast{}^2\right) ^{(n+m)}\Omega\right\|^2. 
\end{align*}
Since 
$$N^l\left( f_NL^\ast{}^2\right) ^{(n+m)}\Omega=
(2(n+m))^l\prod_{j=1}^{n+m}f(2j) 
L^\ast{}^{2(n+m)}\Omega,$$ 
the radius of convergence of the above infinite series is given by 
$\lim_{n\to\infty} n/|f(2n+2m))|= M_f$. Then 
the lemma follows. 
\qed
\begin{lemma}\label{ev_gL0}
Let $f$ and $g$ be complex-valued functions on $\NN$ such that \eqref{fg} is satisfied. 
Then, for all $l, m, n\in\NN$ and $\alpha\in\CC$ such that $|\alpha|<M_f$, 
\begin{align*}
\Omega\in\rD\left( \left(g_{N+2}L^2\right)^l N^m \left(f_NL^\ast{}^2\right)^ne^{\alpha f_NL^\ast{}^2}\right) .
\end{align*}
Moreover, $e^{\alpha f_N{L^\ast}^2}\Omega$ is an eigenvector of 
$g_{N+2}L^2$ such that 
\begin{align}\label{t1}
g_{N+2}L^2 e^{\alpha f_N{L^\ast}^2}\Omega=
\alpha\beta e^{\alpha f_N{L^\ast}^2}\Omega.
\end{align}
\end{lemma}
\proof
We have 
\begin{align*}
\left\|\left(g_{N+2}L^2\right)^l N^m \left(f_NL^\ast{}^2\right)^me^{\alpha f_NL^\ast{}^2}\Omega\right\|^2
=\sum_{n=0}^\infty \frac{|\alpha|^{2n}}{(n!)^2}\left\|\left(g_{N+2}L^2\right)^l N^m\left( f_NL^\ast{}^2\right) ^{(n+m)}\Omega\right\|^2. 
\end{align*}
Since 
\begin{align*}
&\left(g_{N+2}L^2\right)^l N^m\left( f_NL^\ast{}^2\right) ^{(n+m)}\Omega\\
&=
\begin{cases}
\left(\prod_{j=n+m-l+1}^{n+m}g_{2j}\right)
(2(n+m))^m 
\left(\prod_{j=1}^{n+m}f_{2j}\right) L^\ast{}^{2(n+m-l)}\Omega,&n+m\geq l,\\
 0, & n+m<l,
\end{cases}
\end{align*} 
the right-hand side above converges for $\alpha\in\CC$ such that 
$|\alpha|<M_f$. 
From Lemma \ref{crfg}, we can also see that $e^{\alpha f_N{L^\ast}^2}\Omega$ is an eigenvector of 
$g_{N+2}L^2$ and \kak{t1} follows. 
\qed
We define the super coherent vector $\xi_{\alpha,f}$ by
$$\xi_{\alpha,f}=e^{\alpha f_N{L^\ast}^2}\Omega.$$
We set
\begin{align}
\DD_{f,\beta}=\left\{ \alpha\in\CC\ \Big|\ 
|1-\alpha\beta|<1,\ 
|\alpha|<M_f
\right\}.
\end{align}

\begin{lemma}\label{log_gf}
Let $f$ and $g$ be complex-valued functions on $\NN$ such that \eqref{fg} is satisfied. 
Then for all $n\in\NN$ and $\alpha\in \DD_{f,\beta}$, we have 
$\left( f_N{L^\ast}^2\right) ^n\xi_{\alpha,f}\in\rD\left( \log\big(g_{N+2}L^2\big)\right) $ and 
\begin{align*}
\log\left( g_{N+2}L^2\right) \left( f_N{L^\ast}^2\right) ^n\xi_{\alpha,f}
=\left\{\left( {f_NL^\ast}^2+\frac{d}{d\alpha}\right) ^n(\log \alpha\beta)\right\} \xi_{\alpha,f}.
\end{align*}
\end{lemma}
\proof 
By Lemma \ref{ev_gL0} we have $\left( f_N{L^\ast}^2\right) ^n\xi_{\alpha,f}\in \bigcap_{m=0}^\infty\rD\left( \big(1-g_{N+2}L^2\big)^m\right) $. 
From \kak{N} and \kak{M} it follows that 
\begin{align*}
&\log\left( g_{N+2}L^2\right) \left( f_N{L^\ast}^2\right) ^n\xi_{\alpha,f}
=-\sum_{k=1}^K \frac{1}{k}\left( 1-g_{N+2}L^2\right) ^k\left( f_N{L^\ast}^2\right) ^n\xi_{\alpha,f}\\
&\quad=-\sum_{k=1}^K\frac{1}{k}\left\{\left( f_N{L^\ast}^2-\ad_{f_NL^\ast{}^2}\right) ^n\left( 1-g_{N+2}L^2\right) ^{k}\right\}\xi_{\alpha,f}\\
&\quad=-\sum_{k=1}^K \frac{1}{k}\left\{\left( f_N{L^\ast}^2+\frac{d}{d\alpha}\right) ^n\left( 1-\alpha\beta\right) ^{k}\right\}\xi_{\alpha,f}
\to\left\{\left( f_N{L^\ast}^2+\frac{d}{d\alpha}\right) ^n(\log\alpha\beta)\right\}\xi_{\alpha,f}
\end{align*}
 as $K\to\infty$. Then the proof is complete. 
\qed

The next theorem is a generalization of (1) of Lemma \ref{super}. 
\begin{theorem}\label{T1}
Let $f$ and $g$ be complex-valued functions on $\NN$ such that \eqref{fg} is satisfied. 
Then 
\begin{align}\label{maru1}
\left[N,\frac{i}{2} \log\left( g_{N+2}L^2\right) \right]=-i\one
\end{align}
holds on the CCR-domain 
$\mathrm{LH}\left\{ \left( f_N{L^\ast}^2\right) ^n \xi_{\alpha,f}\ \big|\ n\in\NN,\ \alpha\in \DD_{f,\beta}\right\} $.
\end{theorem}
\proof
In this proof we set $X=f_NL^\ast{}^2$ and $Y=g_{N+2}L^2$.
From Lemmas \ref{radc}, \ref{ev_gL0} and \ref{log_gf}, we see that, for all $n\in\NN$, 
\begin{align*}
X^n \xi_{\alpha,f}\in\rD\left( N(\log Y)\right) \cap\rD\left( (\log Y)N\right) .
\end{align*}
Since $\beta N=2\overline{XY}$ on $P_0\rD(N)$,
we have
\begin{align*}
(\log Y)NX^n \xi_{\alpha,f}
&=\frac{2}{\beta}(\log Y)\left( YX-\beta\one\right) X^n \xi_{\alpha,f}\\
&=\frac{2}{\beta}\left( Y(\log Y)X^{n+1}-\beta(\log Y)X^n \right) \xi_{\alpha,f}. \end{align*}
By Lemma \ref{log_gf} we have 
\begin{align*}
(\log Y)NX^n \xi_{\alpha,f}&=\frac{2}{\beta}\left\{ (Y(X-\ad_X)-\beta)(\log Y)X^n \right\} \xi_{\alpha,f}\\
&=\left\{ \left( N-\frac{2}{\beta}Y\ad_X\right) (\log Y)X^n\right\} \xi_{\alpha,f}\\
&=\left\{N(\log Y)X^n-\frac{2}{\beta}Y(X-\ad_X)^n\ad_X(\log Y)\right\}\xi_{\alpha,f}\\
&=\left\{N(\log Y)X^n-\frac{2}{\beta}(X-\ad_X)^nY\ad_X(\log Y)\right\}\xi_{\alpha,f}.
\end{align*}
Here we used $Y(X-\ad_X)Z\varphi=(X-\ad_X)YZ\varphi$ for any linear operator $Z$ and any $\varphi\in\rD(XYZ)\cap\rD(YXZ)\cap\rD(YZX)$. 
In a similar way to \eqref{M}, we can obtain that 
\begin{align*}
(\log Y)NX^n \xi_{\alpha,f}&=\left\{N(\log Y)X^n+\frac{2}{\beta}\left( X+\frac{d}{d\alpha}\right) ^n\alpha\beta\frac{d}{d\alpha}\log\left( \alpha\beta\right) \right\}\xi_{\alpha,f}\\
&=\left( N(\log Y)+2\one\right) X^n\xi_{\alpha,f}.
\end{align*}
Hence $[N,\frac{i}{2}\log Y]X^n \xi_{\alpha,f} =-i X^n\xi_{\alpha,f}$ follows and \kak{maru1} is proven. 
\qed

We can also see a generalization of (2) of Lemma \ref{super}. 
\begin{theorem}\label{T2}
Let $\tilde h$ be a complex-valued function on $\NN$ such that $\tilde h_{N+1}^{-1}$ is bounded and $\tilde f$ and $\tilde g$ be complex-valued functions on $\NN $ such that 
\begin{align*}
\tilde h_{N+1}^{-1}\tilde g_{N+3}\tilde h_{N+3}\tilde f_{N+2}-\tilde h_{N-1}^{-1}\tilde g_{N+1}\tilde h_{N+1}\tilde f_NP_{\geq2}
=
\tilde\beta\one\quad\end{align*}
on $P_0\ell_\mathrm{fin}^2$ with some $\tilde\beta\in\CC$. 
Let 
\begin{align}
\DD_{\tilde f, \tilde h, \tilde\beta}=\left\{\alpha\in\CC\ \Bigg|\ 
\alpha\in \DD_{\tilde f,\tilde\beta},\ 
\xi_{\alpha,\tilde f}\in 
\bigcap_{n=0}^\infty\rD\left( \tilde g_{N+3}\tilde h_{N+3}\left( \tilde f_N{L^\ast}^2\right) ^n \right) \right\}.
\end{align}
 Then $\tilde h_NL^\ast \xi_{\alpha,\tilde f}$ is an eigenvector of 
$\tilde g_{N+2}L^2 $ such that 
\begin{align}\label{ab}
\tilde g_{N+2}L^2 \tilde h_NL^\ast \xi_{\alpha,\tilde f}&=\alpha\tilde\beta \tilde h_NL^\ast \xi_{\alpha,\tilde f}.
\end{align}
Furthermore
it follows that 
\begin{align}\label{maru2}
\left[N,\frac{i}{2} \log\left( \tilde g_{N+2}L^2\right) \right]&=-i\one
\end{align}
on the CCR-domain 
$ 
\mathrm{LH}\left\{ \tilde h_NL^\ast \left( \tilde f_N{L^\ast}^2\right) ^n\xi_{\alpha,\tilde f}\ \middle|\ n\in\NN,\ \alpha\in \DD_{\tilde f, \tilde h, \tilde\beta}\right\} $.
\end{theorem}
\proof 
By the following equalities 
\begin{align*}
\tilde g_{N+2}L^2\tilde h_NL^\ast&=\tilde g_{N+2}\tilde h_{N+2}L
=\tilde h_N\tilde h_N^{-1}\tilde g_{N+2}\tilde h_{N+2}L^\ast L^2
=\tilde h_NL^\ast\left( \tilde h_{N+1}^{-1}\tilde g_{N+3}\tilde h_{N+3}L^2\right) 
\end{align*}
on $P_0\eln\cap\rD(\tilde g_{N+1}\tilde h_{N+1})$, 
one can show \kak{ab} in a similar manner to the proof of \kak{fg} by 
replacing $g_{N+2}$ of \eqref{fg} with $\tilde h_{N+1}^{-1}\tilde g_{N+3}\tilde h_{N+3}$. 
The proof of \kak{maru2} is also similar to those of Lemma \ref{ev_gL0} 
and Theorem \ref{T1}. 
\qed

By Theorems \ref{T1} and \ref{T2} we can also construct an ultra-weak time operator of $N$. This is a generarization of $\cT$ in \kak{T6}. 
Let $f$, $\tilde f$, $g$, $\tilde g$ and $h$ be functions on $\NN$ 
given 
in Theorems~\ref{T1} and \ref{T2}. 
Let
\begin{align*}
&\cK_0=\mathrm{LH}\left\{\left( f_N{L^\ast}^2\right) ^n\xi_{\alpha,f}\ \Big|\ n\in\NN,\ \alpha\in \DD_{f,\beta}\right\},\\
&\cK_1=\mathrm{LH}\left\{\tilde h_NL^\ast\left( \tilde f_N{L^\ast}^2\right) ^n 
\xi_{\alpha,\tilde f}\ \Big|\ n\in\NN,\ \alpha\in \DD_{\tilde f, \tilde h, \tilde\beta}\right\}.
\end{align*}
Note that 
$\cK_0\oplus
\cK_1$
is dense if $\DD_{f,\beta}\neq\emptyset$ and $\DD_{\tilde f, \tilde h, \tilde\beta}\neq\emptyset$. 
Let 
\begin{align*}
&S_{0g}=\frac{i}{2} \log\left( g_{N+2}L^2\right) ,\\
&S_{1\tilde g}=\frac{i}{2} \log\left( \tilde g_{N+2}L^2\right) . 
\end{align*}
These are generalization of $S_0$ and $S_1$. 
Define 
\begin {align*}
&\cT_{0g}[\varphi,\psi]=\half\{(S_{0g}\varphi,\psi)+(\varphi,S_{0g}\psi)\},
\quad \varphi,\psi\in \cK_0,\\
&\cT_{1\tilde g}[\varphi,\psi]=\half\{(S_{1\tilde g} \varphi,\psi)+(\varphi,S_{1\tilde g}\psi)\},\quad 
\varphi,\psi\in \cK_1
\end{align*}
and set 
\begin{align}
\label{uwto}
\cT_{g,\tilde g}=\cT_{0g}\oplus \cT_{1\tilde g}.
\end{align}
\begin{theorem}
Suppose that $f,\tilde f, g, \tilde g$ and $\tilde h$ satisfy 
the assumptions given in Theorem \ref{T1} and Lemma \ref{T2}. 
We also assume that $\DD_{f,\beta}\neq\emptyset$ and $\DD_{\tilde f, \tilde h, \tilde\beta}\neq\emptyset$.
Then $\cT_{g,\tilde g}$ is an ultra-weak time operator of $N$ 
with 
the dense CCR-domain 
$\cK_0\oplus
\cK_1$. 
\end{theorem}
\proof This follows from \kak{maru1} and \kak{maru2}. 
\qed

\subsection{Conjugate operators of the form \textbf{$\bm{\log p(L)$}}}
\label{44}
In the previous section we consider the conjugate operator of the form 
$\log (g_NL^2)$. 
Another direction of generalizations is to consider 
conjugate operators of the form 
$\log (g_NL^n)$. 
In special cases 
it can be reduced to considering  
conjugate operators of 
$\log (g_NL^n)$ as those of 
$\log (L^n)$. See Lemma \ref{T7} below. 
Let $f$ be a function on $\NN$ and 
\begin{align*}f!_k(N)\xi_n=\left(\prod_{m=0}^{[n/k]} f_{n-km}\right)\xi_n,\quad n\in\NN.
\end{align*}
We extend $f!_k(N)$ to $\ell_{\rm fin}^2$ by the linearity and 
denote the closure of $f!_k(N)\lceil_{ \ell_{\rm fin}^2}$ by 
the same symbol $f!_k(N)$. 
\begin{lemma}\label{T7}
Let $f$ and $g$ be complex-valued functions on $\NN$ such that 
$f_ng_n=n$ for all $n\in\NN$. 
Suppose that $\alpha\in(0,1)$ and 
$\Omega\in\rD(e^{\alpha f_NL^\ast{}^k})$. 
Then 
\begin{align}
\label{te3}\log\left(g_{N+k}L^k\right)e^{\alpha f_N{L^\ast}^{k}}\Omega=f!_k(N)\log\left((N+k\one)L^k\right)e^{\alpha {L^\ast}^k}\Omega.
\end{align}
In particular, setting $h_n=n$ for all $n\in\NN$, we have 
\begin{align}
\label{te1}&\left(\log L^k\right)e^{\alpha N{L^\ast}^{k}}\Omega=
h!_k(N)\log\left((N+k\one)L^k\right)e^{\alpha {L^\ast}^k}\Omega,\\
\label{te2}&\log\left(g_{N+k}L^k\right)e^{\alpha f_N{L^\ast}^k}\Omega=
h!_k(N)\f f!_k(N)\left(\log L^k\right)e^{\alpha N{L^\ast}^{k}}\Omega. 
\end{align}
\end{lemma}
\proof
It is straightforward to see that 
\begin{align*}
\log\left(g_{N+k}L^k\right)e^{\alpha f_N{L^\ast}^{k}}\Omega&=\log\left(g_{N+k}L^k\right)f!_k(N)e^{\alpha {L^\ast}^{k}}\Omega\\
&=f!_k(N)\log\left(g_{N+k}f_{N+k}L^k\right)e^{\alpha {L^\ast}^k}\Omega\\
&=f!_k(N)\log\left((N+k\one)L^k\right)e^{\alpha {L^\ast}^k}\Omega.
\end{align*}
Putting $g_n=1$ for all $n\in\NN$ in \kak{te3}, 
we can derive \kak{te1}. 
Combining \kak{te3} and \kak{te1} we can also see \kak{te2}. 
 Then the lemma follows. 
\qed
By \kak{te2}, 
$\log\big(g_{N+k}L^k\big)$ can be represented 
as 
$C(N) \log L^k$ on $\mathrm{LH} \big\{e^{\alpha N{L^\ast}^{k}}\Omega\big\}$ with 
\[C(N)=h!_k(N)\f f!_k(N).\] 
Since the operator $C(N)$ commutes with $N$, 
the investigation of conjugate operators of the form $\log\big(g_{N+k}L^k\big)$
 can be reduced to that of 
$\log L^k$ on 
$\mathrm{LH}\big\{L^\ast{}^{nk}
e^{\alpha N{L^\ast}^k}\Omega \ \big|\ n\in\NN,\ |\alpha|<1\big\}$.
\begin{example}\label{LLA}
We see that 
$\TA\cong \LA\oplus \LLA$ with 
$\LA=
\frac{i}{2}\log\left( \sqrt{\frac{N+2\one}{N+\one}}L^2\right)$ and 
$\LLA=\frac{i}{2}\log\left( \sqrt{\frac{ N+\one}{{N+2\one}}}L^2\right)$.   
By Lemma \ref{T7}, the investigation of the CCR-domain of $\LA$ and $\LLA$ can be reduced to 
investigating the CCR-domain of $\frac{i}{2}\log L^2 $. 
\end{example}

Thus from now on we study the conjugate operator of the form 
$\log p(L)$.  
The approach to finding a CCR-domain for 
$\log p(L)$ is similar to that of the previous subsection. 
Formally 
\begin{equation*}
[N, \log p(L)]\subset -p(L)\f p'(L) L, 
\end{equation*}
and $e^{\alpha N L^\ast}\Omega$ is an eigenvector of both $L$ and $p(L)$. 
Thus 
$[N, \log p(L)]=-p(\alpha)\f p'(\alpha) \alpha$ on 
$\mathrm{LH}\{ e^{\alpha N L^\ast}\Omega\}$. 
To solve $[N, \log p(L)]=c$,  
we study the algebraic equation of the form 
\begin{align*} \alpha p'(\alpha)+cp(\alpha)=0\end{align*}
for some $c\in\CC$. 
Then 
we define 
\begin{align*}
&\rD(p,c)\\
&=\mathrm{LH}\{\varphi\in \rD(N)\mid \exists \alpha\in \CC\ s.t. \ |p(\alpha)-1|<1,\ \alpha p'(\alpha)+cp(\alpha)=0,\ \varphi\in\ker(L-\alpha\one)\}.
\end{align*}
It is possible  that $\rD(p,c)=\{0\}$. 
\begin{theorem}\label{T3}
Let $p$ be a polynomial and $c\in\CC\setminus\{0\}$. 
If there exists $\alpha\in\CC$ such that $|\alpha|<1$, 
$\alpha p'(\alpha)+cp(\alpha)=0$ and $|p(\alpha)-1|<1$, 
then $\rD(p,c)\not=\{0\}$ and 
\begin{align*}
\left[N, -\frac{i}{c}\log p(L)\right]=-i\one
\end{align*}
with the CCR-domain $\rD(p,c)$. 
Moreover, in $\rD\big(\sum k^{-1}N(\one-p(L))^k\big)$, there is no infinite dimensional CCR-domain for $N$ and $c\log p(L)$ unless $p(x)=cx^m$ for some $c\in\CC $ and $m\in\NN$.
\end{theorem}
\proof
We can see that 
\begin{align}\label{[N,logp(L)]}
 [N, \log p(L)]=-\sum_{k=1}^\infty\frac{1}{k}\left[N,\left( \one-p(L)\right) ^k\right]
 =-\sum_{k=1}^\infty\left( \one-p(L)\right) ^{k-1}Lp'(L)
 \end{align}
on $\rD(\sum k^{-1}N(\one-p(L))^k)\cap\rD(N\log p(L))\cap\rD(\log p(L)N)$. Let $\varphi\in\ker(L-\alpha\one)$.
Then 
\begin{align*}
 [N, \log p(L)]\varphi=-\alpha p'(\alpha)p(\alpha)^{-1}\varphi.
\end{align*}
By $\alpha p'(\alpha)+cp(\alpha)=0$ we see that 
\begin{align*}
 [N, \log p(L)]\varphi
 =c\varphi.
\end{align*}
Next we shall show that the dimension of the CCR-domain in $\ran(p(L))$ is finite. 
For any 
\begin{equation*}
\varphi\in\rD\left(\sum_{k\geq1}\frac{1}{k}N(\one-p(L))^k\right)
\end{equation*}
 in the CCR-domain, there exists 
$c\in\CC$ such that $[N, \log p(L)]\varphi=c\varphi$. From 
 \eqref{[N,logp(L)]}, we see that $\lim_{k\to\infty}(\one-p(L))^kLp'(L)\varphi=0$. Therefore, 
\begin{align*}
cp(L)\varphi=p(L)[N, \log p(L)]\varphi
=-\sum_{k=1}^\infty p(L)\left( \one-p(L)\right) ^{k-1}Lp'(L)\varphi
=-Lp'(L)\varphi.
\end{align*}
Thus $\varphi\in\ker(Lp'(L)+cp(L))$. 
This implies that the CCR-domain is at most finite dimensional space, 
since $\dim\ker(L-\lambda\one)^k\leq k$ for all $\lambda\in \CC$ and $k\in\NN$, and $p$ is a polynomial with $Lp'(L)+cp(L)\neq0$.
This fact can be derived from a fundamental lemma below. 
Then the theorem follows. 
\qed

\begin{lemma}
Let $A$ be a bounded operator on a Banach space $\cK$. 
Let 
$\{n_1,\ldots,n_k\}\subset\NN$,
$\{\lambda_1,\ldots,\lambda_k\}\subset \CC$ such that
$\lambda_i\neq\lambda_j$ for $i\neq j$.
Then it follows that 
\begin{align*}\ker \left( \prod_{j=1}^k (\la_j\one-A)^{n_j}\right) =
\bigoplus_{j=1}^k \ker (\la_j\one-A)^{n_j}.\end{align*}
\end{lemma}
\proof
We refer to \cite[Lemma 1.76]{Aie04}.
\qed

\section{\textbf{Galapon operators}}
\label{4}
\subsection{Galapon operators by shift operators}
In this section we investigate Galapon operator which is 
a bounded self-adjoint time operator. 
Recall that $\s(N)=\{n\}_{n\in\NN}$ 
and $\{v_n\}_{n\in\NN}$ is the set of normalized eigenvectors of $N$ in $L^2(\mathbb R)$. 
We define $\TG$ by 
\begin{align}
&\rD(\TG)=\mathrm{LH}\{v_n\mid n\in\NN\},\nonumber\\
&\TG \varphi=i\sum_{n=0}^\infty\left( \sum_{m\neq n}\frac{(v_m,\varphi) }{n-m}v_n\right),\quad \varphi\in\rD(\TG) \label{G}.
\end{align}
It is known that $\TG$ is bounded and 
\begin{align}\label{galapon}
[N, \TG]=-i\one
\end{align}
on the  dense CCR-domain $\mathrm{LH}\{v_n-v_m\mid n, m\in\NN\}$ 
(see \cite{AM08}). 
In particular, $\TG$ is not equal to the angle operator $\TA$. 
The angle operator 
$\TA$ is expressed on $\eln$ by Theorem \ref{47}. 
 On the other hand the Galapon operator is given by \kak{G}. 
As it stands, no direct comparison of $\TA$ and $\TG$ can be made, 
so we shall transform $\TG$ into an operator on $\eln$. 
Let us consider $\log\left( \one-L\right)$ and $\log\left( \one-L^\ast\right)$. 

\begin{lemma}\label{212}
We have
$\ell_\mathrm{fin}^2\subset \rD\left( \log(\one-L)\right)\cap\rD\left( \log(\one-L^\ast)\right) $. 
\end{lemma}
\proof
By the definition of $\log(\one-L) $, we see that 
\begin{align*}\log(\one-L)\xi_n=-\sum_{k=1}^\infty \frac{1}{k}L^k \xi_n=-\sum_{k=1}^n \frac{1}{k}\xi_{n-k}.\end{align*} 
Thus $\xi_n\in \rD\left( \log(\one-L)\right) $ and hence $\ell_\mathrm{fin}^2\subset \rD(\log(\one-L))$. 
We also see that 
\begin{align*}\log(\one-L^\ast)\xi_n=-\sum_{k=1}^\infty \frac{1}{k}{L^\ast}^k \xi_n=-\sum_{k=1}^\infty \frac{1}{k}\xi_{n+k}\end{align*}
and 
\begin{align*}
\left\|\log(\one-L^\ast)\xi_n\right\|^2=\sum_{k=1}^\infty \frac{1}{k^2}<\infty. 
\end{align*}
Thus $\xi_n\in \rD\left( \log(\one-L^\ast)\right) $ and $\ell_\mathrm{fin}^2\subset \rD\left( \log(\one-L^\ast)\right) $. 
\qed
\begin{remark}
By the proof of Lemma \ref{212} we see that 
\bi
\item[{\rm (1)}] $\ell_\mathrm{fin}^2\subset\rD(\log(\one-L^\ast))$, but 
$\rD(\log L^\ast)\cap \ell_\mathrm{fin}^2=\{0\}$, 
\item[{\rm (2)}] $\log (\one-L)\ell_\mathrm{fin}^2\subset \ell_\mathrm{fin}^2$. 
\ei
\end{remark}

We consider the following operator $L_G$.
\begin{definition}
We define a linear operator $L_G$ on $\ell^2$ by 
\begin{align*}
&\rD(L_G)=\left\{\varphi\in\ell^2\ \middle|\ \lim_{K\to\infty}\sum_{k=1}^K \frac{1}{k}\left( L^\ast{}^k- L^k\right) \varphi\text{ exists}\right\},\\
&L_G\varphi=i\sum_{k=1}^\infty \frac{1}{k}
\left( L^\ast{}^k- L^k\right) \varphi,\quad \varphi\in\rD(L_G).
\end{align*}
\end{definition}

Note that $\ell^2_\mathrm{fin}\subset\rD(L_G)$ and $i\{\log(\one-L)-\log(\one-L^\ast)\}\subset \LG$.

\begin{theorem}\label{UT}
\begin{enumerate}[{\rm (1)}]
\item It holds that $L_G\cong T_G$ on $\ell^2_\mathrm{fin}$.
\item $\LG$ is a bounded operator.
\end{enumerate}
\end{theorem}
\proof
(1) Let $U$ be the unitary operator defined by $Uv_n=\xi_n$ for each $n\in\NN$. 
For arbitrary $\varphi\in\rD(T_G)$, we see that 
\begin{align*}
UT_G\varphi
&
=i\sum_{n=0}^\infty\left( \sum_{m<n}\frac{\left( \xi_m, U\varphi\right) }{n-m}+\sum_{m>n}\frac{\left( \xi_m, U\varphi\right) }{n-m}\right) \xi_n\\
&=i\sum_{n=0}^\infty\left( \sum_{m<n}\frac{\left( L^{n-m}\xi_n, U\varphi\right) }{n-m}-\sum_{m>n}\frac{\left( L^\ast{}^{m-n}\xi_n, U\varphi\right) }{m-n}\right) \xi_n\\
&=i\sum_{n=0}^\infty\left( \sum_{k=1}^\infty\frac{\left( L^k \xi_n, U\varphi\right) }{k}-\sum_{k=1}^\infty\frac{\left( L^\ast{}^k \xi_n, U\varphi\right) }{k}\right) \xi_n.
\end{align*}
From Lemma \ref{212}, we have $U\rD(T_G)=\ell^2_\mathrm{fin}\subset\rD(L_G)$. 
Then 
\begin{align*}
UT_G\varphi &=i\sum_{n=0}^\infty
\left( 
\sum_{k=1}^\infty\frac{1}{k}\left( 
L^k-L^\ast{}^k\right) \xi_n, U\varphi\right)\xi_n
=\sum_{n=0}^\infty
\left( \xi_n, i\sum_{k=1}^\infty\frac{1}{k}\left( L^\ast{}^k- L^k\right) U\varphi\right) \xi_n.
\end{align*}
This implies that $UT_G\varphi=L_G U\varphi$ for $\varphi\in\rD(T_G)$. 

(2) From the Hilbert inequality, 
we see that, for any $\varphi=\sum_n c_n\xi_n\in\rD(L_G)$,
\begin{equation*}
|(\varphi, L_G\varphi)|=\left|\sum_{n=0}^\infty\sum_{m\neq n}\frac{c_nc_m}{n-m}\right|\leq\pi\|\varphi\|^2.
\end{equation*}
Thus we have $\|L_G\varphi\|\leq\pi\|\varphi\|$. 
Then the theorem is proven. 
\qed


\begin{theorem}\label{NLG}
We have
$[N, \LG]=-i\one$ on 
$\rD(N\LG )\cap \rD(\LG N)$.
\end{theorem}
\proof
We obtain that, for any $\varphi\in\rD(N\LG )\cap \rD(\LG N)$,
\begin{align}
[N,L_G]\varphi
&=\sum_{n=0}^\infty\left( \xi_n,(N\LG -\LG N)\varphi\right) \xi_n\nonumber \\
&=i\sum_{n=0}^\infty\sum_{k=1}^\infty\frac{1}{k}\left( \xi_n,\left(\left[N,{L^\ast}^k\right]-\left[N,L^k\right]\right)\varphi\right) \xi_n \nonumber \\
&\label{cor}=i\sum_{n=0}^\infty\sum_{k=1}^\infty
\left( \xi_n,\left( {L^\ast}^k+L^k\right) \varphi\right) \xi_n.
\end{align}
Let $\varphi=\sum_{n=0}^\infty c_n\xi_n$. 
Since 
$
\sum_{n=0}^\infty\sum_{k=1}^\infty
\left( \xi_n,\left( {L^\ast}^k+L^k\right) \varphi\right) \xi_n
=
\sum_{n=0}^\infty(\sum_{m\neq n}c_m)\xi_n
$, we have 
$$\infty>\|[N,\LG]\varphi\|^2=
\sum_{n=0}^\infty\left|\left(\sum_{m\neq n}c_m\right)\right|^2.$$
In particular 
$$
\left\|
\sum_{n=0}^\infty\left(\sum_{m\neq n}c_m\right)\xi_n
+\varphi\right\|^2
=
\sum_{n=0}^\infty
\left|\sum_{m=0}^\infty c_m\right|^2<\infty,$$
which implies that 
$\varphi\in\big\{\sum_{n=0}^\infty c_n\xi_n\in\ell^2\ \big|\ \sum_{n=0}^\infty c_n=0\big\}$, and 
\begin{equation*}
\sum_{n=0}^\infty\sum_{k=1}^\infty\left( \xi_n,\left( {L^\ast}^k+L^k\right) \varphi\right) \xi_n=-\varphi.
\end{equation*}
Hence $[N,L_G]\varphi=-i\varphi$ and the theorem is proven. 
\qed
It is immediate to see that 
$[N,\TG]\phi=-i\phi$ does not hold true for $\phi=v_n$, $n\in\NN$. 
It is however shown in \cite{gal02a} that a CCR-domain of $\TG$ is 
$\mathrm{LH}\{v_n-v_m\mid n, m\in\NN\}$. 
Note that 
$$U\left( \mathrm{LH}\{v_n-v_m\mid n, m\in\NN\}\right)=
(\one-L^\ast)\ell^2_\mathrm{fin}.$$ 
Hence 
$[N, \LG]=-i\one$ holds on 
$(\one-L^\ast)\ell^2_\mathrm{fin}$ and 
$(\one-L^\ast)\ell^2_\mathrm{fin}\subset 
\rD(N\LG )\cap \rD(\LG N)$.
\begin{remark}
To our knowledge, it was previously unknown that
 $[N, \LG]=-i\one$ holds on a strictly larger subspace 
than $(\one-L^\ast)\ell^2_\mathrm{fin}$. 
It can be shown however that the CCR-domain 
$\rD(N\LG )\cap \rD(\LG N)$ of $\LG$ 
is strictly larger 
than $(\one-L^\ast)\ell^2_\mathrm{fin}$. 
It can be actually seen that 
$(\one-L^\ast)e^{\alpha L^\ast}\Omega
\in(\rD(L_GN)\cap\rD(NL_G))\setminus(\one-L^\ast)\ell^2_\mathrm{fin}$. 
\end{remark}
\begin{remark}\label{T9}
In \eqref{cor}, 
if $\varphi\in\ran(\one-L^\ast)$, 
then we have
\begin{align}
[N,L_G]\varphi=i\sum_{n=0}^\infty\sum_{k=1}^\infty\left( \xi_n,\left( {L^\ast}^k+L^k\right) \varphi\right) \xi_n
\label{cor1}=
i\left((\one-L^\ast)^{-1}+(\one-L)^{-1}-2\one\right)\varphi. 
\end{align}
\kak{cor1} can be extended in \kak{cor2}. 
\end{remark}

\subsection{\textbf{Generalization of Galapon operator $\bm{\TG}$}}
In Theorem \ref{UT} we show that 
$\TG$ can be expressed as 
$i\{\log(\one-L)-\log(\one-L^\ast)\}$ on $\ell^2_\mathrm{fin}$. 
We shall generalize this. 
We set 
\begin{align*}L_g=i\left\{ 
\log\left( \one-g_{N}L\right) -\log\left( \one-L^\ast g_{N}^{-1}\right) \right\}.\end{align*}

\begin{lemma}\label{uni}
Let $g$ be a complex-valued function on $\NN $ such that
$|g_n|=1$ for all $n\in\NN$. 
Then $L_g$ is unitary equivalent to $L_G$ on $\ell^2_\mathrm{fin}$.
\end{lemma}
\proof
We can construct the unitary operator $V$ on $\eln $ such that
$g_{N}L=V^\ast LV$. 
It is actually given by 
\begin{align*}
V(c_0,c_1,c_2,c_3,\ldots,)=
(c_0,g_0c_1,g_0g_1 c_2, g_0g_1g_2c_3,\ldots,).
\end{align*}
Since $|g_n|=1$, $V$ is unitary. 
Thus $L_g$ is unitary equivalent to $L_G$.
\qed

Let $\ell_1^2(g)=\big\{\sum_{n=0}^\infty c_n\xi_n\in\eln\ \big|\  \sum_{n=0}^\infty g!(n)c_n=0\big\}$, where $g!(n)=\prod_{k\leq n}g_k$.
\begin{theorem}\label{N_Lg}
Let $g$ be a complex-valued function on $\NN$.
\begin{enumerate}[{\rm (1)}]
\item Suppose that 
$|g_n|>0$ for all $n\in\NN$. 
Then 
$\left[N, L_g\right]=-i\one$ 
 on 
$\rD(NL_g)\cap\rD(L_gN)$. \\
\item Suppose that 
$|g_n|=1$ for all $n\in\NN$. 
Then $L_g$
is a bounded time operator with the CCR-domain
$\rD(NL_g)\cap\rD(L_gN)$. 
\end{enumerate} 
\end{theorem}
\proof 
As in the proof of Theorem \ref{NLG}, we see that 
$\rD(NL_g)\cap\rD(L_gN)\subset \ell_1^2(g)$.
For any $\varphi\in\rD(NL_g)\cap\rD(L_gN)$, we have 
\begin{align*}
[N,L_g]\varphi
=i\sum_{n=0}^\infty\sum_{k=1}^\infty
\left( \xi_n,\left( \left( L^\ast g_{N}^{-1}\right) ^k+\left( g_{N}L\right) ^k\right) \varphi\right) \xi_n
=-i\varphi.
\end{align*}
Then (1) is proven. 
Since $\LG$ is self-adjoint and bounded, 
$\overline{L_g}$ is also self-adjoint and bounded from Lemma~\ref{uni}. 
By (1) $L_g$ satisfies 
the canonical commutation relation. 
\qed
$L_g$ is no longer a symmetric operator when $|g|\neq1$. 
Therefore, in order to construct a time operator from $L_g$ with 
$|g|\neq1$, we need to symmetrize it. 
\begin{corollary}
Let $g$ be a complex-valued function on $\NN $. 
Suppose that there exist $n_0\in\NN$, $c_1,\ c_2\in(0,\infty)$ and $d_1,\ d_2\in(0,1/2)$ such that, for all $n>n_0$, 
\begin{align}\label{con_g}
c_1 n^{-d_1}\leq \prod_{k=0}^n |g_k|\leq c_2 n^{d_2}.
\end{align}
Then $(L_g+L_{\bar g^{-1}})/2$ is a time operator of $N$. 
\end{corollary}
\proof
The assumption $c_1 n^{-d_1}\leq \prod_{k\leq n} |g_k|$ implies that for any $m\in\mathbb N$ 
\begin{align*}
\left\|\sum_{k\geq1}\frac{1}{k}\left(L^\ast g_N^{-1}\right)^k\xi_m\right\|
&=\left\|\sum_{k\geq1}\frac{1}{k}\left(\prod_{l=0}^{k-1}g_{m+l}\right)^{-1}\!\!\xi_{m+k}\right\|\\
&\leq\left(\sum_{1\leq k\leq n_0}\!\!\!\!\frac{1}{k^2}\left|\prod_{l=0}^{k-1}g_{m+l}\right|^{-2}\!\!+\left|c_1^{-1}g!(m-1)\right|^2\sum_{k>n_0}k^{-2(1-d_1)}\right)^{1/2}<\infty.
\end{align*}
Thus $\xi_m\in \rD(\log\left( \one-L^\ast g_{N}^{-1}\right))$ and 
$\xi_m\in \rD(L_g)$. 
In the same way, 
$\prod_{k=0}^n |g_k|\leq c_2 n^{d_2}$ implies that 
$\xi_n\in \rD(\log\left( \one-L^\ast \bar g_{N}\right))$
and then
$\xi_n\in \rD(L_{\bar g^{-1}})$. 
Thus \kak{con_g} implies that $\xi_n\in \rD(L_g)\cap \rD(L_{\bar g^{-1}})$. 
Since 
$L_g^\ast \supset -i \left\{ \log(\one-L^\ast \bar g_N)-\log(\one-\bar g_N^{-1} L)\right\}=L_{\bar g^{-1}}$,
the operator $L_g+L_{\bar g^{-1}}$ is symmetric. 
The canonical commutation relation
$\big[N, (L_g+L_{\bar g^{-1}})/2\big]=-i\one$ 
 is proven 
in the same way as Theorem \ref{N_Lg}.
\qed

\subsection{\textbf{Conjugate operators of the form 
$\bm{\log (\one-p(L))-\log (\one-p(L^\ast))}$}}
In Section \ref{44} conjugate operators of 
the form $\log p(L)$ is considered. 
Let 
$Y_p=\one-p(L)$. 
In this section 
we consider conjugate operators of the form 
\begin{align*}
X_p=\log(Y_p)-\log(Y_p^\ast),
\end{align*}
under some conditions on the polynomials $p$. 
$X_p$ is a generalization of $\LG$. 
Note that $\ker(Y_p^\ast)=\{0\}$ if $p\neq1$. 
Therefore 
the inverse of $Y_p^\ast$ exists. 
\begin{lemma}\label{tera1}
Let $p$ be a polynomial. 
Suppose that $Y_p$ is injective and $\lim_{k\to\infty}p(L)^k=0$. 
Then 
\begin{align}\label{l1}
\left[N,\log Y_p \right]=L p'(L)Y_p^{-1}
\end{align}
on $\ran(Y_p)\cap \rD\left( N\log Y_p \right) \cap \rD\left( \log(Y_p)N\right)$ 
and
\begin{align}\label{l2}
\left[N,\log(Y_p^\ast)\right]=-L^\ast p'(L^\ast)(Y_p^\ast)^{-1}
\end{align}
on $\ran(Y_p^\ast)\cap \rD\left( N\log(Y_p^\ast) \right) \cap \rD\left( \log(Y_p^\ast) N\right) $.
\end{lemma}
\proof 
Let $\varphi\in \rD\left( N\log Y_p \right) \cap \rD\left( \log(Y_p) N\right)$ and 
$\varphi= Y_p \psi$.
Inserting the expansion $\varphi=\sum_{n=0}^\infty (\xi_n,\varphi)\xi_n$ to 
$[N,\log Y_p ]\varphi$, 
we see that 
\begin{align*}
\left[N,\log Y_p \right]\varphi
=-\sum_{n=0}^\infty\sum_{k=1}^\infty \frac{1}{k}\left( \xi_n, \left[N,p(L)^k\right]\varphi\right) \xi_n.
\end{align*}
Since $[N,p(L)^k]\subset -kLp'(L)p(L)^{k-1}$, we have
\begin{align*}
\left[N,\log Y_p \right]\varphi&=\sum_{n=0}^\infty\sum_{k=1}^\infty \left( \xi_n, L p'(L)p(L)^{k-1}\varphi\right) \xi_n\\
&=\sum_{n=0}^\infty\sum_{k=1}^\infty \left( \xi_n, L p'(L)p(L)^{k-1}(\one-p(L))\psi\right) \xi_n. 
\end{align*}
By assumption $\lim_{k\to\infty}p(L)^k=0$, we have 
\begin{align*}
\left[N,\log Y_p \right]\varphi
=-\sum_{n=0}^\infty\lim_{k\to\infty}\left( \xi_n, Lp'(L)p(L)^k\psi\right) \xi_n+\sum_{n=0}^\infty\left( \xi_n, Lp'(L)\psi\right) \xi_n
=Lp'(L)Y_p ^{-1}\varphi.
\end{align*}
Then \kak{l1} follows. \kak{l2} can be shown in a similar way to \kak{l1}. 
\qed

From Lemma \ref{tera1}, if $Lp'(L)Y_p^{-1}+L^\ast p'(L^\ast)(Y_p^\ast)^{-1}$ has a non-zero eigenvalue, then $N$ has 
a conjugate operator on the eigenvector space. 
Let $\SS^1=\{\alpha\in\CC\mid |\alpha|=1\}$. 
Thus, we have the following theorem.
\begin{theorem}\label{N_fL}
Let $p$ be a polynomial with degree $m$ such that 
\begin{align*}
&(1)\ \one-p(L) \mbox{ is injective},\\
&(2)\ \lim_{k\to\infty}p(L)^k=0,\\
&(3)\ \{z\in\CC\mid 1-p(z)=0\}\subset \SS^1.
\end{align*}
Let 
$\alpha_1,\ldots,\alpha_m\in\CC$ be all roots of $1-p(\alpha)=0$ 
within the multiplicity. 
Then 
\begin{align*}
\left[N, \frac{i}{m} X_p\right]=-i\one
\end{align*}
with the CCR-domain 
$\ran\big(\prod_{i=1}^m(\alpha_i L^\ast-\one)\big)\cap\rD(X_pN)\cap \rD(NX_p)$.
\end{theorem}
\proof
By Lemma \ref{tera1} we have 
\begin{align}
[N, X_p]&=
Lp'(L)Y_p^{-1}+L^\ast p'(L^\ast)(Y_p^\ast)^{-1}
\nonumber\\
&\label{cor2}=\sum_{i=1}^m \left( \alpha_i(\alpha_i\one-L)^{-1}+\alpha_i(\alpha_i\one-L^\ast)^{-1}-2\one\right) .
\end{align}
Note that \kak{cor2} corresponds to \kak{cor1}. 
It suffices to show that 
the right-hand side above has an eigenvector in 
$\rD\left( NX_p\right) \cap \rD\left( X_pN\right) $. 
Since $\alpha_i\in\SS^1$,
we see that 
\begin{align*}
 \sum_{i=1}^m \left( \alpha_i(\alpha_i\one-L)^{-1}+\alpha_i(\alpha_i\one-L^\ast)^{-1}\right) =\sum_{i=1}^m \left( \alpha_i(\alpha_i\one-L)^{-1}+\alpha_i^{-1}(\alpha_i^{-1}\one-L^\ast)^{-1}\right) .
\end{align*}
Let $\psi\in \eln \setminus\{0\}$ and $
\varphi=(\alpha L^\ast-\one) \psi$. 
Since
\begin{align*}
\left( \alpha(\alpha\one-L)^{-1}+\alpha^{-1}(\alpha^{-1}\one-L^\ast)^{-1}\right) \varphi&=\alpha L^\ast \psi-\psi=\varphi,
\end{align*}
$\varphi$ is an eigenvector of 
$\alpha(\alpha\one-L)^{-1}+\alpha^{-1}(\alpha^{-1}\one-L^\ast)^{-1}$ 
corresponding to the  eigenvalue $1$. 
Hence
$\prod_{i=1}^m(\alpha_i L^\ast-\one)\psi$ is an eigenvector of 
$Lp'(L)Y_p^{-1}+L^\ast p'(L^\ast)(Y_p^\ast)^{-1}$ corresponding to  the eigenvalue $-1$ and 
\begin{equation*}
[N,X_p]\prod_{i=1}^m(\alpha_i L^\ast-\one)\psi=-m \prod_{i=1}^m(\alpha_i L^\ast-\one)\psi.
\end{equation*}
Then the CCR-domain is given by 
$\ran(\prod_{i=1}^m(\alpha_i L^\ast-\one))\cap\rD(X_pN)\cap \rD(NX_p)$, 
and the theorem is proven. 
\qed
\begin{corollary}\label{T8}
Let $|\om|=1$. Then $im^{-1}\{\log(\one-\om L^m)-\log(\one-\om^\ast L^\ast{}^m)\}$ is a time operator of $N$. 
\end{corollary}
\proof
Let $p(z)=\omega z^m$. Then the corollary follows from Theorem \ref{N_fL}.
\qed

\section{Classification of conjugate operators}
\label{5}
In this section we focus on investigating conjugate operators of the form 
\[T_{\om,m}=\frac{i}{m}\log(\om\one-L^m)\quad (\om,m)\in\overline \DD\times(\NN\setminus\{0\}),\]
which have appeared in Lemma \ref{212} and  Corollary \ref{T8}. 
Furthermore
note that 
$\TA\cong \LA\oplus \LLA$. 
In Example \ref{LLA}
we mentioned that 
investigating the CCR-domain of $\LA$ and $\LLA$ can be reduced to investigating the CCR-domain of $$\frac{i}{2}\log L^2=T_{0,2}.$$ 
Let us define the family of operators with parameters $\om$ and $m$ by
$$\sT=\{T_{\om,m}\mid \om\in\overline \DD,m\geq1\}.$$
We divide $\sT$ into three disjoint families:
\begin{align*}\sT=\sT_{\{0\}}\cup\sT_{\DD\setminus\{0\}}\cup\sT_{\partial \DD}.
\end{align*}
Here 
\begin{align*}
&\sT_{\{0\}}=\{T_{\om,m}\mid \om=0,m\geq1\},\\
&\sT_{\DD\setminus\{0\}}=\{T_{\om,m}\mid 0< |\om|<1,m\geq1\},\\
&\sT_{\partial \DD}=\{T_{\om,m}\mid |\om|=1,m\geq1\}. 
\end{align*}
In what follows we discuss CCR-domains of $T_{\om,m}$.

We define
\begin{align*}\ell_m^2=\left\{\varphi\in\eln\ \middle|\ \lim_{k\to\infty}(\one-L^m)^k\varphi=0\right\},\end{align*}
\begin{theorem}[$\sT_{\{0\}}$]\label{Y1}
We have
\[\left[N, \frac{i}{m}\log L^m\right]=-i\one\]
on $\ell_m^2\cap\mathrm \rD(N\log L^m)\cap\mathrm \rD(\log(L^m)N)$, 
and 
$\dim \left( \ell_m^2\cap\mathrm \rD(N\log L^m)\cap\mathrm \rD(\log(L^m)N)\right)=\infty$. 
\end{theorem}
\proof
This is a special case of Theorem \ref{T3}. 
Let $p(z)=z^m$. Then 
$zp'(z)-mp(z)=0$ for all $z\in\CC$. 
In a similar way of the proof of Theorem \ref{T3}, for any $\varphi\in
\ell_m^2\cap\mathrm \rD(N\log L^m)\cap\mathrm \rD(\log(L^m)N)$, 
we see that 
$[N, \log L^m]\varphi
=-m\sum_{k=1}^\infty\left( \one-L^m\right) ^{k-1}L^m\varphi=-m\varphi$.
Then the theorem follows. 
\qed

Next we consider $\sT_{\DD\setminus\{0\}}$. 
Let $c\in\CC$ and we recall that 
\begin{align*}
&\rD(\om-z^m,c)\\
&=\mathrm{LH}\{\varphi\mid \exists \alpha\in \CC\ s.t. \ |1-\om+\alpha^m|<1,\ \alpha^m=c(m+c)^{-1}\om,\ \varphi\in\ker(L-\alpha\one)\}.
\end{align*}
\begin{theorem}
[$\sT_{\DD\setminus\{0\}}$]\label{Y2}
Let $\om\in \DD\setminus\{0\}$. 
Then for $c\in \CC\setminus\{0\}$ such that 
$|c|$ is sufficiently small, we see that 
$\rD(\om-z^m,c)\neq\{0\}$, $\dim \rD(\om-z^m,c)<\infty$ 
and \begin{align*}
\left[N, -\frac{i}{c}\log(\om\one-L^m)\right]=-i\one
\end{align*}
on $\rD(\om-z^m,c)$. 
Moreover, for any $c\in \CC \setminus\{0\}$, there is no infinite dimensional CCR-domain for $N$ and $c\log(\om-L^m)$
\end{theorem}
\proof
Let $p(z)=\om-z^m$ with $|\om|<1$ and $|\om-1|<1$. Suppose that $c$ is a sufficiently small positive number. 
Then $z p'(z)+cp(z)=-(c+m)z^m+c\om=0$ has 
the roots 
\[\alpha_k=\left|\frac{c}{c+m}\right|\om^{1/m} e^{2\pi ik/m},\quad
k=0,\ldots, m-1.\]
By Theorem \ref{T3}, 
since $|p(\alpha_k)-1|= |\om-1-\alpha_k^m|<1$ and $|\alpha_k|<1$, 
$k=0,\ldots, m-1$, we have 
$[N, \frac{i}{c}\log(\om\one-L^m)]=i\one$ on 
$\ker(L-\alpha_k\one)$ for $k=0,\ldots, m-1$.
Moreover the dimension of the CCR-domain 
$\mathrm{LH}\{\ker(L-\alpha_k\one), k=0,\ldots, m-1\}$ is finite.
\qed
We consider conjugate operators of the form 
$i\{\log(\omega\one-L)-\log(\bar\omega \one-L^\ast)\}$ with 
$\om\in\partial\DD$. 
We also consider $\Log (\om\one-A)$ 
for 
$\log (\om\one-A)$ in order to avoid a singularity. 
\begin{definition}[$\Log (\om\one-A)$]
\label{DEF}
Let $A$ be a linear operator on a Hilbert space $\cH$. 
We define
\begin{align}
&\rD(\Log (\om\one-A))=\left\{\varphi\in \bigcap_{k=0}^\infty\rD(A^k)\ \middle|\ 
\lim_{K\to\infty}\sum_{k=1}^K \frac{1}{k}\left(\frac{1}{\om}A\right)^k\!\!\!\varphi\text{ exists} \right\},\nonumber\\
&\Log (\om\one-A)\varphi=\log(\om)\varphi-\sum_{k=1}^\infty \frac{1}{k}\left(\frac{1}{\om}A\right)^k\!\!\!\varphi,\quad \varphi\in\rD(\Log(\om\one-A))\label{log w_1}.
\end{align}
\end{definition}
\begin{theorem}
[$\sT_{\partial \DD}$]\label{Y3}
Let $\om\in\partial\DD$.
Then $L_{m,\om}=im^{-1}(\Log(\om\one-L^m)-
\Log(\bar\om\one-L^\ast{}^m))$
is a time operator with the dense CCR-domain
$\rD(NL_{m,\om})\cap\rD(L_{m,\om}N)$. 
\end{theorem}
\proof
As in the proof of Theorem \ref{NLG}, we see that 
$$\rD(NL_{m,\om})\cap\rD(L_{m,\om}N)\subset
\left.
\left\{ 
\sum_{n=0}^\infty c_n\xi_n\right|
\sum_{n=0}^\infty\bar\omega^nc_{l+mn}=0\text{ for all }l< m\right\}$$ and 
for any $\varphi\in\rD(NL_{m,\om})\cap\rD(L_{m,\om}N)$, 
\begin{align*}
[N, L_{m,\omega}]\varphi&=i\sum_{n=0}^\infty \sum_{k=1}^\infty\left(\xi_n, \left((\bar\omega L^m)^k+(\omega L^\ast{}^m)^k\right)\varphi\right)\xi_n=-i\varphi.
\end{align*}
Then the theorem follows. 
\qed
We summarize results obtained in Theorems \ref{Y1}, \ref{Y2} and \ref{Y3} 
in Table \ref{t3}.  
\begin{table}[h]
\centering
\renewcommand{\arraystretch}{1.5}
\begin{tabular}{l|c|c|c} 
 & $\sT_{\{0\}}$ 
 & $\sT_{\DD\setminus\{0\}}$
 & $\sT_{\partial \DD}$\\
 \hline
$T_{\om,m}$ & $\om=0$ 
 & $\om\in\DD\setminus\{0\}$
 & $\om\in\partial\DD$\\
 \hline
boundedness & unbounded 
 & unbounded
 & bounded\\
 \hline
CCR-domain& infinite dim. 
 & finite dim.
 & dense\\
 \hline
example & $\TA$ 
 & 
 & $\TG=T_{1,1}+T_{1,1}^\ast$\\
 \hline
$\log$ & Def \ref{def} 
 & Def \ref{def}
 & Def \ref{DEF}
 \end{tabular}
\caption{Classification of $T_{\om,m}$}
\label{t3}
\end{table}

We can also introduce operator $T_{\om,m}$ with $|\om|>1$ 
by a Dunford integral. 
See \kak{ses} in Appendix. 

\section{Weak Weyl relations for $\bm N$}
\label{6}
In this section we consider the time evolution of conjugate operators. 
The Weyl relation \cite{wey28} is 
$$ e^{-isp}e^{-itq}=e^{ist}e^{-itq}e^{-isp},\quad s,t\in\RR.$$
From this we can derive
the so-called weak Weyl relation \cite{miy01}:
\begin{align}\label{1111}
q e^{-it p}=e^{-itp} (q+t),\quad t\in\RR,
\end{align}
on $\rD(q)$. 
The strong time operator $T$ is defined as an operator satisfying the above weak Weyl relation \kak{1111} with
$q$ and $p$ replaced by $T$ and the Hamiltonian $H$, respectively. More precisely,
we explain the weak Weyl relation \kak{1111} as follows.
\begin{definition}
\label{im1}
We say that the pair $(H,A)$ consisting of a self-adjoint operator $H$ and a symmetric operator $A$ on a Hilbert space $\mathcal H$ 
obeys the weak Weyl relation if for all $t\in\RR$,
\bi
\item[(1)]
 $e^{-itH}\rD(A)\subset \rD(A)$;
 \item[(2)] $Ae^{-itH}\varphi=e^{-itH}(A+t)\varphi$ for all $\varphi\in \rD(A)$.
\ei
\end{definition}
Here $A$ is referred to as a strong time operator associated with $H$. 
Note that a strong time operator is not unique.
If strong time operator $A$ is self-adjoint, then
it is known that
$$
e^{-isA}e^{-itH}=e^{-ist} e^{-itH}e^{-isA},\quad s,t\in\RR
$$
holds.
In particular when Hilbert space is separable, by the von Neumann uniqueness theorem
 the Weyl relation above implies that $H$ and $A$ are unitarily equivalent to
 $\oplus^n p$ and $\oplus^n q$ with some $n\leq\infty$, respectively.
As mentioned above,
 although a strong time operator is automatically
a time operator,
the converse is not true.
It is remarkable that when the pair $(H,A)$ obeys
the weak Weyl relation, $H$ has purely absolutely continuous spectrum.
For example since the spectrum of $N$ is purely discrete, 
there is no strong time operator associated with $N$. 
Weak Weyl relation can be understood as the time evolution of $A$: 
\begin{align}\label{wwr2}
e^{itH}Ae^{-itH}\supset A+t
\end{align}
We can see the time evolution of the Galapon operator directly. 
It can be seen that 
$$\TG(t)\varphi=i\sum_{n=0}^\infty\sum_{m\neq n}
\frac{e^{+it(n-m)}(v_m,\varphi)}{n-m}v_n.$$
Hence $\TG(t)$ is periodic with period $2\pi$ so that 
$$\TG(t+2\pi n)=\TG(t),\quad n\in\NN, t\in\RR.$$
Let us consider the time evolutions of general conjugate operators. 
We define 
$$T_{\om,m}(t)=
e^{itN}T_{\om,m}e^{-itN}.$$
\begin{theorem}
Let $T_{\om,m}\in\sT$. 
Then 
$$ T_{\om,m}(t)=
\frac{i}{m}\log (\om-e^{-itm} L^m)$$
and 
$T_{\om,m}(t)$ is periodic with 
period $2\pi/m$, i.e., 
$$T_{\om,m}(t)=T_{\om,m}(t+(2\pi/m)n),\quad n\in\NN,t\in\RR.$$
\end{theorem}
\proof
Since $e^{itN}Le^{-itN}=e^{-it}L$, 
we see that $e^{itN}\log(\om\one-L^m)e^{-itN}=
\log (\om-e^{-itm}L^m)$. 
\qed

\begin{example}
It is shown that 
the angle operator $\TA=\half(\arctan q\f p\oplus\arctan pq\f)$ can be represented as 
\begin{align*}
&\half\arctan \left(q\f p\right)\cong 
\frac{i}{2}\log\left( \sqrt{\frac{N+2\one}{N+\one}}L^2\right),\\
&\half \arctan \left(pq\f\right)\cong \frac{i}{2}\log\left( \sqrt{\frac{ N+\one}{{N+2\one}}}L^2\right)
\end{align*}
in Theorem \ref{ang}. 
Thus the period of the time evolutions of 
both $\arctan\left(q\f p\right)$ and 
$\arctan\left(pq\f\right)$ is $\pi$. 
\end{example}

Finally we show an example of \kak{log2}. 
Let $\om=0$ and consider the time evolution of $T_{0,m}(t)$. 
\begin{corollary}
Let $m\in\NN$ and $\arg e^{-itm }\neq 0$. 
Then 
\begin{align*}
\log (-e^{-itm }L^m)
\neq \log\left(e^{-itm }\right)+\log (-L^m).
\end{align*}
\end{corollary}
\proof
Suppose that 
$\log (-e^{-itm }L^m)
=\log\left(e^{-itm}\right)+\log (-L^m)
$. 
Then we see that 
\begin{align*}
e^{itN} T_{0,m}e^{-itN}&=
\frac{i}{m}\log (-e^{-itm }L^m)=
-\frac{1}{m}\arg e^{-itm }+\frac{i}{m}\log (-L^m)
=
-\frac{1}{m}\arg e^{-itm }+T_{0,m}.
\end{align*}
Hence for any eigenvector $v_n$ of $N$ we have
\begin{align}\label{cot}
(v_n, T_{0,m}v_n)=(v_n , e^{itN} T_{0,m}e^{-itN}v_n)=
 -\frac{1}{m}(v_n, \arg e^{-itm }v_n)+(v_n, T_{0,m}v_n).
 \end{align}
 Since $(v_n, \arg e^{-itm }v_n)\neq 0$, \kak{cot} leads to a contradiction. 
 Then the corollary follows. 
\qed

\appendix
\counterwithin*{equation}{section} 
\renewcommand\theequation{\thesection.\arabic{equation}} 
\section{CCR domains for $\sT_{\ov{\DD}^c}$}
We consider conjugate operators of the form 
$i\log(\om\one-L^m)$ with $|\om|>1$. 
To avoid singularities, 
we need to modify the definition of $i\log(\om\one-L^m)$. 
Let $r\DD=\{rz\mid z\in \DD\}$ for $r>0$ 
and $f$ be an analytic function such that $f(\ov{r\DD})\not\ni0$ for some $r>1$. Then $\log f$ is analytic on $r\DD$. 
Let $\|A\|<1$. Since $(z-A)\f$ is bounded for any $z\not \in \ov{\DD}$, 
the Dunford integral 
$\oint_{r\partial \DD}\log f(z) (z-A) dz$ 
defines the sesquilinear form 
\begin{align}
\label{ses}
(\varphi,\psi)\mapsto 
Q(\varphi,\psi)=\oint_{r\partial \DD}\log f(z) (\varphi, (z-A)\f \psi) dz. 
\end{align}
Equation \kak{ses} also defines the bounded operator $B$ such that 
$Q(\varphi,\psi)=(\varphi, B\psi)$. 
We also denote $B$  by 
$\log f(A)$, following the notation of Definition~\ref{def}.
By choosing $f(z)=\omega-z^m$ and $A=L$, we can define 
$\log(\omega\one-L^m)$ by \kak{ses}. 
\begin{lemma}\label{T5}
There is no polynomial $p$ satisfying the following conditions:
\begin{enumerate}[{\rm (1)}]
\item $0\not\in\s \left( p(L)\right) $, 
\item $[N, i\{\log p(L)-\log p(L^\ast)\}]=-i\one$ on some infinite dimenstional subspace
$\cD$. 
\end{enumerate}
\end{lemma}
\proof Let $p(x)=\sum_{k=0}^n a_kx^k$ be a real-valued polynomial with $a_n\neq0$. 
We see that 
\begin{equation}\label{p1}
p(L)p(L^\ast){L^\ast}^n=\left( \sum_{k=0}^na_kL^\ast{}^{n-k}\right) \left( \sum_{k=0}^na_k{L^\ast}^{k}\right) 
=\sum_{k,l\geq0}^n a_ka_{l}L^\ast{}^{n-l+k}.
\end{equation}
It is easy to see that 
\begin{equation}\label{N_logp}
[N, \log p(L)-\log p(L^\ast)]\subset-Lp'(L)p(L)\f -L^\ast p'(L^\ast)p(L^\ast)\f.
\end{equation}
Observe that  $\ran({L^\ast}^n)$ is given by 
$\ran({L^\ast}^n)=\{(c_k)_{k\in\mathbb N}\in\eln \mid 
c_0=\cdots=c_{n-1}=0\}$. Hence $\cD\cap 
\ran({L^\ast}^n)=\{0\}$ implies that $\dim\cD<\infty$. 
However 
$\dim\cD=\infty$ by hypothesis, 
we must have 
$\cD\cap\ran({L^\ast}^n)\neq\{0\}$. 
Notice that, for any $k\leq n$, $L^k{L^\ast}^k={L^\ast}^kL^k=\one$ on 
$\ran({L^\ast}^n)$. 
Thus we also have the following relation on the non-trivial domain by 
 \eqref{N_logp}:
\begin{align}
&p(L)p(L^\ast){L^\ast}^n\nonumber\\
&=-p(L)p(L^\ast)[N, \log p(L)-\log p(L^\ast)]{L^\ast}^n\nonumber\\
&=Lp'(L)p(L^\ast){L^\ast}^n+L^\ast p'(L^\ast)p(L){L^\ast}^n\nonumber\\
&=\left( \sum_{k=1}^n ka_kL^\ast{}^{n-k}\right) \left( \sum_{k=0}^na_k{L^\ast}^{k}\right) +\left( \sum_{k=1}^nka_k{L^\ast}^{k}\right) \left( \sum_{k=0}^na_kL^\ast{}^{n-k}\right) \nonumber\\
&=\sum_{k,l\geq 1}^n (ka_ka_{l}+la_{l}a_k)L^\ast{}^{n-l+k}
+\sum_{k=1}^n ka_ka_0L^\ast{}^{n-k}
+\sum_{k=1}^n ka_ka_0L^\ast{}^{n+k}.\label{p2}
\end{align}

Let $L^\ast{}^n\varphi\in\rD\setminus\{0\}$ and $n_0=\inf\mathop{\mathrm{supp}}(\varphi)$. 
Comparing 
the coefficients of $L^\ast{}^0$ in $\eqref{p1}$ and $\eqref{p2}$, 
we have
$na_0a_n=a_0a_n$. 
From (1) we have $a_0\neq0$. 
Since $a_0a_n\neq 0$, we see that $n=1$. 
Hence $p(x)=ax+b$. 
By \kak{p2} we also have $a=\pm b$, 
and then $p(x)=a(1\pm x)$. 
We conclude that 
$p(L)=a(\one\pm L)$ and $\s(p(L))=p(\s(L))\ni 0$. 
This contradicts $0\not\in\s \left( p(L)\right)$.
Then the lemma follows. 
\qed
\begin{theorem}
Let $|\om|>1$. The operator $i\{\log(\omega\one-L^m)-\log(\bar{\omega}\one-L^\ast{}^m)\}$ has no infinite dimensional CCR-domain for $N$. 
\end{theorem}
\proof
Let $p(z)=\om-z^m$ with $|\om|>1$.
Then $\s(p(L))\not\ni 0$. Thus the theorem follows from Lemma \ref{T5}. 
\qed
\begin{remark}\label{apprem}
The existence of a non zero CCR domain of 
$i\{\log(\omega\one-L)-\log(\bar\omega \one-L^\ast)\}$ with $|\omega|>1$ 
is unknown. 
\end{remark}

{\bf Acknowledgements:}
FH is financially supported by  
JSPS KAKENHI 20H01808,
 JSPS KAKENHI 20K20886,JSPS KAKENHI 23K20217 and 
 JSPS KAKENHI 25H00595.

\bibliographystyle{plain}{\small \bibliography{hiro7}}

\end{document}